\pdfoutput=1
\documentclass[11pt,twoside,a4paper,cmspaper,final,collab]{cms-tdr}
\def\svnVersion{5716f1b}\def\svnDate{2026/04/07}\def\cmsCernNoTag{CERN-EP-2025-210}\def\cmsCernDate{\today}\def\cmsMessage{Published in Physics Letters B as \href{http://dx.doi.org/10.1016/j.physletb.2026.140359}{\doi{10.1016/j.physletb.2026.140359}.}}
\begin{document}\cmsNoteHeader{HIN-24-004}

\newcommand{\PbPb}{\ensuremath{\text{PbPb}}\xspace}
\newcommand{\XeXe}{\ensuremath{\text{XeXe}}\xspace}
\renewcommand{\Pb}{\ensuremath{\text{Pb}}\xspace}
\newcommand{\Xe}{\ensuremath{\text{Xe}}\xspace}
\providecommand{\MUSIC}{\textsc{music}\xspace}
\ifthenelse{\boolean{cms@external}}
{
  \newcommand{\cmsLeft}{upper\xspace}
  \newcommand{\cmsRight}{lower\xspace}
}
{
  \newcommand{\cmsLeft}{left\xspace}
  \newcommand{\cmsRight}{right\xspace}
}
\cmsNoteHeader{HIN-24-004}

\title{Characterizing the initial state and dynamical evolution in \texorpdfstring{\XeXe}{XeXe} and \texorpdfstring{\PbPb}{PbPb} collisions using multiparticle cumulants}

\author*[cern]{Aryaa Dattamunsi}
\author*[cern]{Prabhat Pujahari}
\author*[cern]{Wei Li}

\date{\today}

\abstract{
For the first time, correlations among mixed-order moments of two or three flow harmonics ---($v_{n}^{k},v_{m}^{l}$) and ($v_{n}^{k},v_{m}^{l}, v_{p}^{q}$), with $k$, $l$, and $q$ denoting the respective orders---are measured in xenon--xenon (\XeXe) collisions and compared with lead--lead (\PbPb) results, providing a novel probe of collective behavior in heavy ion collisions. These measurements compare a nearly spherical, doubly-magic ${}^{208}$\Pb nucleus to a triaxially deformed ${}^{129}$\Xe nucleus, emphasizing the sensitivity to initial-state geometry fluctuations arising from nuclear deformation. The dependence of these results ($v_{n}$, $n = 2, 3, 4$) on the shape and size of the nuclear overlap region is studied. Comparisons between $v_{2}$, $v_{3}$, and $v_{4}$ demonstrate the importance of $v_{3}$ and $v_{4}$ in exploring the nonlinear hydrodynamic response of the quark-gluon plasma (QGP) to the initial spatial anisotropy. The results constrain initial-state model parameters that influence the evolution of the QGP. The CMS detector was used to collect \XeXe and \PbPb data at nucleon-nucleon center-of-mass energies of $\sqrtsNN = 5.44$ and 5.36\TeV, respectively. Correlations are extracted using multiparticle mixed-harmonic cumulants (up to eight-particle cumulants) with charged particles in the pseudorapidity range $\abs{\eta} < 2.4$ and transverse momentum range $0.5 < \pt < 3$\GeV/c.
}

\hypersetup{%
pdftitle={Characterizing the initial state and dynamical evolution in XeXe and PbPb collisions using multiparticle cumulants},%
pdfauthor={CMS Collaboration},%
pdfsubject={CMS},%
pdfkeywords={CMS, heavy ion, flow, QGP, system size}} 

\maketitle

\section{Introduction}

High energy heavy ion collisions at the LHC create a hot and dense state of strongly interacting 
matter, known as the quark--gluon plasma (QGP)~\cite{BRAHMS:2004adc, cms2025overview}. The anisotropic shape of the 
overlap region between the two colliding nuclei generates pressure gradients that drive a collective 
anisotropic expansion of the medium. This collective behavior, known as anisotropic flow~\cite{Ollitrault:1992bk}, is quantified 
through a Fourier decomposition of the single-particle azimuthal distribution of produced particles~\cite{Voloshin:1994mz}:
\begin{equation}\label{eq:vn} 
\frac{dN}{d\phi} = \frac{1}{2\pi} \left[1 + 2\sum_{n=1}^{\infty} v_n \cos n(\phi - \psi_n) \right], 
\end{equation}
where $\phi$ is the azimuthal angle of an outgoing particle, $\psi_n$ denotes the $n$th-harmonic symmetry plane, 
and $v_n$ is the $n$th flow harmonic. The coefficients $v_2$, $v_3$, and $v_4$ correspond to elliptic, triangular, 
and quadrangular flow, respectively. These harmonics carry information about the initial spatial anisotropy 
and the subsequent medium response.

Event-by-event fluctuations in the positions of nucleons within the colliding nuclei lead 
to variations in the initial energy density, causing flow harmonics to fluctuate on an event-by-event 
basis. These fluctuations are studied through higher-order cumulants of the flow coefficients 
($v_{n}^{k}$, $k>1$). While second-order cumulants of a single flow harmonic and two-harmonic 
correlations are well explored~\cite{CMS:2017kcs, ALICE:2019zfl}, they are insensitive to 
non-Gaussian features of the flow probability distribution~\cite{Noronha-Hostler:2015dbi}. 
The study of mixed-order cumulants 
(correlations between different harmonics and orders, $v_{n}^{k}$ and $v_{m}^{l}$, with 
$n \neq m$ and $k \neq l$) 
provides sensitivity to the nonlinear hydrodynamic response of the QGP. 
A purely linear response results in $v_{n}$ scaling linearly with the corresponding 
initial-state eccentricity $\varepsilon_{n}$, which quantifies the transverse shape anisotropy 
of the collision zone immediately after impact~\cite{Noronha-Hostler:2015dbi}. 

In addition to event-by-event fluctuations in the energy density within each nucleus, 
the shape of the colliding nuclei, whether spherical or deformed, also influences the 
initial geometry of the overlap region and, consequently, the observed flow. 
The doubly magic lead ($^{208}$\Pb) nuclei are nearly spherical \cite{Bender:2003jk}, 
whereas xenon ($^{129}$\Xe) nuclei are triaxially deformed \cite{Bally:2021qys}. 
This study therefore probes differences in the initial-state fluctuations arising from 
event-by-event variations in the nucleon configuration in collisions of 
nuclei with different shapes. The nuclear shapes are characterized by 
static ground-state deformation parameters constrained by low-energy nuclear physics 
measurements~\cite{Jia:2021tzt}. Following the parameterization used in 
low-energy nuclear structure studies, 
the nuclear density distribution is described using the Woods--Saxon formula:

\begin{equation}\label{eq:rho}
\rho(r, \theta) = \frac{\rho_0}{1 + \exp[(r - R(\theta)) / a_0]},
\end{equation}
where $r$ is the radial distance from the center of the nucleus, 
$\theta$ is the polar angle relative to the symmetry axis, $\rho_0$ is the central density at $r = 0$, 
and $a_0$ is the nuclear skin depth. For a deformed nucleus, the radial term $R(\theta)$ includes 
angular dependence via spherical harmonics:
\begin{equation}\label{eq:Rtheta}
R(\theta) = R_0 \left[1 + \beta_2 Y_{20}(\theta) + \beta_4 Y_{40}(\theta) \right],
\end{equation}
where $R_0$ is the mean nuclear radius, $\beta_2$ and $\beta_4$ are the quadrupole and hexadecapole 
deformation parameters, respectively, and $Y_{\ell 0}(\theta)$ are the spherical harmonics.

This Letter presents a comprehensive study of multiparticle cumulants involving one, two, and 
three flow harmonics, including mixed higher-order correlations between 
different harmonics. The analysis is based on data collected with the CMS detector for \XeXe 
and \PbPb collisions at nucleon-nucleon center-of-mass energies of $\sqrtsNN = 5.44$ and 5.36\TeV, 
respectively. The \XeXe and \PbPb data were recorded in 2017 and 2023, respectively, and correspond 
to integrated luminosities of 3.42\mubinv~\cite{LUM-17-003,LUM-17-004,LUM-18-002} and 1.92\nbinv. The key goals of this 
study are: (i) to investigate the effects of nuclear deformation on flow correlations using \XeXe 
and \PbPb comparisons, and (ii) to probe the nonlinear hydrodynamic response of the QGP using 
higher-order moments of $v_{n}$. The results are compared to a range of theoretical models to constrain the 
initial-state geometry and the transport properties of the QGP formed in heavy ion collisions at the 
LHC. Tabulated results for this analysis can be accessed in the HEPdata record~\cite{HEPData}. 

\section{The CMS detector}

The CMS detector is centered around a superconducting solenoid with an internal diameter of 6\unit{m}, generating a 3.8\unit{T} magnetic 
field~\cite{CMS:2014pgm, CMS:2008xjf}. Within the solenoid volume reside a silicon pixel and strip tracker, a lead tungstate 
crystal electromagnetic calorimeter (ECAL), and a brass-and-scintillator hadron calorimeter (HCAL), each composed of a barrel 
and two endcap sections. Forward calorimeters extend the pseudorapidity ($\eta$) coverage of the barrel and endcap detectors. Muons are 
identified using gas-ionization chambers embedded in the steel flux-return yoke outside the solenoid. The hadron forward (HF) 
calorimeters, positioned symmetrically 11.2\unit{m} from the interaction region, cover the range $3.0<\abs{\eta}<5.2$ and 
also serve as luminosity monitors. These HF calorimeters use steel absorbers and quartz fibers as sensitive material, are azimuthally 
segmented into $20^{\circ}$ modular wedges, and form towers of size $0.175 \times 0.175$ in ($\Delta\eta \times \Delta\phi$). 
The silicon tracker, consisting of 1856 silicon pixel and 15\,148 silicon strip detector modules, measures charged particles within 
$\abs{\eta}<2.4$. For nonisolated particles with transverse momentum $1<\pt<10$ \GeV/c and $\abs{\eta} < 1.4$, the typical track momentum resolution is 
1.5\%, and the transverse (longitudinal) impact parameter resolution ranges from 25--90 (45--150) \mum. The CMS detector response 
is simulated in detail using \GEANTfour~\cite{GEANT4:2002zbu}.

\section{Event and track selections}

Events from \XeXe collisions are selected at the hardware level (level-1) by requiring the presence of both colliding bunches at the nominal interaction point~\cite{CMS:2019cyz}. A minimum energy deposit is required in at least one HF calorimeter tower, optimized to maximize event acceptance while suppressing noise from electromagnetic scattering and pileup (multiple interactions in a single bunch crossing). Pileup is negligible because the average number of collisions per bunch crossing is 0.018. At the high-level trigger~\cite{TRG-12-001}, at least one reconstructed track in the pixel detector is required. Offline event selection further requires a minimum energy deposit of 3\GeV in each of at least three HF towers on both sides of the detector, to enhance selection of hadronic collisions and suppress ultraperipheral electromagnetic interactions. Events must contain a reconstructed primary vertex with at least two tracks, and located within 15\unit{cm} along the beam axis and 0.2\unit{cm} transversely relative to the nominal collision region. To reduce beam-gas interaction contamination, events with more than ten tracks must have at least 25\% of tracks meeting high-purity quality criteria~\cite{CMS:2014pgm}. The overall event selection efficiency for hadronic events is approximately 95\%. Track reconstruction follows algorithms similar to those used in proton-proton collisions~\cite{CMS:2014pgm}.

For \PbPb collisions, a cluster shape filter in the pixel detector ensures compatibility with particles originating from the primary vertex. The pileup rate is negligible (on average 0.0005 collisions per bunch crossing). Events are selected by requiring energy deposits of at least 4\GeV in two HF towers on both sides of the interaction point, favoring hadronic collisions. The same vertex location criteria as in \XeXe data are used. Track reconstruction for the \PbPb data utilizes a deep neural network algorithm trained with input variables including normalized $\chi^{2}$ per degree of freedom, \pt resolution, and number of hits in the silicon pixel and strip detectors, based on conditions of the CMS detector from the reconstruction of earlier 2018 \PbPb data~\cite{CMS:2019cyz}.

In both the \XeXe and \PbPb systems, the analysis is restricted to primary charged particle tracks with $\abs{\eta} < 2.4$ and $0.5 < \pt < 3.0$ \GeV/c. For \XeXe data, tracks are required to have impact parameter significances with respect to the primary vertex less than three standard deviations in both the beam direction ($d_{z}$) and the transverse plane ($d_{0}$), as well as a relative \pt uncertainty $\sigma_{\pt}/\pt < 10\%$. For \PbPb data, the neural-network-based reconstruction imposes requirements on $d_{z}$ and $d_{0}$ similar to those used for \XeXe data \cite{CMS:2014pgm, Giannini:2023jpr}.

In this Letter, all observables are presented as functions of centrality, which is related to the degree of geometric overlap of the two colliding nuclei. The centrality percentage represents the fraction of the total inelastic hadronic cross section. A 0\% centrality corresponds to a fully head-on collision, while higher centrality percentages indicate more peripheral collisions with smaller overlap. Experimentally, centrality is determined using the percentiles of the energy deposited in the HF calorimeters~\cite{CMS:2012krf}.

\section{Analysis techniques}

The two-particle correlation and multiparticle cumulant methods are discussed in Ref.~\cite{CMS:2019cyz}. 
In the two-particle correlation analyses, a charged particle from one \pt interval (the "trigger" particle) 
is paired with all of the remaining charged particles from the same \pt interval (the "associated" particles).

For a given trigger particle, the pairing is performed in two-dimensional (2D) bins of differences in $\eta$ and azimuthal angle 
$(\Delta\eta,\Delta\phi)$. The signal and mixed-event distributions are defined as the per-trigger-particle yield of pairs in the same or mixed events:
\begin{equation}\label{eq:eq4}
\begin{aligned}
S(\Delta\eta,\Delta\phi) &= \frac{1}{N_{\text{trig}}} 
\frac{\rd^{2}N^{\text{same}}}{\rd\Delta\eta\, \rd\Delta\phi}\\
M(\Delta\eta,\Delta\phi) &= \frac{1}{N_{\text{trig}}} 
\frac{\rd^{2}N^{\text{mixed}}}{\rd\Delta\eta\, \rd\Delta\phi},
\end{aligned}
\end{equation}
where $N^{\text{same}}$ ($N^{\text{mixed}}$) is the number of same- (mixed-) event particle pairs in a given $(\Delta\eta,\Delta\phi)$ bin and $N_{\text{trig}}$ is 
the number of trigger particles.

The mixed-event distribution is used to estimate the random combinatorial background and is 
constructed by pairing trigger particles in an event with associated particles from ten other randomly 
chosen events. Events are considered for mixing only if they belong to the same centrality bin (with a bin width of 5\% or 10\%) and the difference in their primary vertex locations along the beam axis is less than 2\unit{cm}.

The per-trigger-particle associated yield used to extract the $v_{n}$ coefficients is found by normalizing the ratio of the signal and mixed-event distributions:
\begin{equation}\label{eq:eq6}
\frac{1}{N_\text{{trig}}}\frac{\rd^{2}N^{\text{pair}}}{\rd\Delta\eta \rd\Delta\phi} = M(0,0) \frac{S(\Delta\eta,\Delta\phi)}{M(\Delta\eta,\Delta\phi)}, 
\end{equation}
where $M(0,0)$ is the mixed-event value at $(\Delta\eta,\Delta\phi) = (0, 0)$. 
This 2D normalized yield is then projected onto $\Delta\phi$ before fitting to a Fourier expansion. 
A selection of $\abs{\Delta\eta}> 2$ is applied to suppress short-range nonflow correlations such as jets~\cite{ALICE:2021adw}. The fitting is done using: 
\begin{equation}\label{eq:eq7}
\frac{1}{N_\text{trig}}\frac{\rd{}N^{\text{pair}}}{\rd\Delta\phi} = \frac{N_\text{assoc}}{2\pi} \left( 1 + \sum_{n} 2 V_{n\Delta} \cos(n \Delta\phi) \right),
\end{equation}
where $N_\text{assoc}$ is the number of associated particles in the selected kinematic interval, and $V_{n\Delta}$ denotes the $n$th Fourier coefficient.

Under the factorization assumption~\cite{CMS:2016fnw}, $V_{n\Delta}(\text{p}_T^a, \text{p}_T^b) = v_n(\text{p}_T^a)\,v_n(\text{p}_T^b)$, 
and for trigger and associated particles from the same \pt range, the $n$th flow harmonic is 
extracted as:
\begin{equation}
v_{n}\{2, \abs{\Delta\eta} > 2\} = \sqrt{V_{n\Delta}},
\end{equation}
where the first "2" in the $v_{n}$ variable list refers to flow harmonics from two-particle correlations. 

Multiparticle correlations involving four or more particles are studied with the \textit{Q}-cumulant 
method~\cite{Bilandzic:2010jr}, which incorporates track-by-track weights to correct for detector acceptance and
efficiency~\cite{Bilandzic:2013kga}. This method allows the computation of cumulants that probe genuine multiparticle
azimuthal correlations. Here, ``genuine'' correlations refer to irreducible multiparticle correlations
that cannot be factorized into products of lower-order correlations, thereby providing sensitivity
to collective behavior while suppressing nonflow contributions.

Within this framework, the four-particle $v_n$, $v_{n}\{4\}$, are extracted from four-particle cumulants.
The two-particle cumulant $c_{n}\{2\}$ and four-particle cumulant $c_{n}\{4\}$ are defined as:
\begin{equation}
c_n\{2\} = \langle\!\langle e^{in(\phi_1 - \phi_2)} \rangle\!\rangle,
\,  
c_n\{4\} = \langle\!\langle e^{in(\phi_1 + \phi_2 - \phi_3 - \phi_4)} \rangle\!\rangle
- 2\,c_n\{2\}^2,
\end{equation}
where $\langle \langle \cdots \rangle \rangle$ denotes an average performed in two steps: 
(i) first over all particles within a single event, and (ii) then over all events~\cite{Bilandzic:2010jr}. 
The corresponding four-particle cumulant is then obtained as
\begin{equation}
v_n\{4\} = \sqrt[4]{-c_n\{4\}} .
\end{equation}

Four-particle symmetric cumulants (SC) measure correlations between the squared flow coefficients of two different harmonics, $n$ and $m$:
\ifthenelse{\boolean{cms@external}}
{
{
  \tolerance=1000
  \emergencystretch=2em
  \begin{multline}
    \mathrm{SC}(n,m) =
    \langle\!\langle
      \cos(n\phi_1 + m\phi_2 - n\phi_3 - m\phi_4)
    \rangle\!\rangle \\
    - \langle\!\langle \cos[n(\phi_1 - \phi_2)] \rangle\!\rangle
      \langle\!\langle \cos[m(\phi_1 - \phi_2)] \rangle\!\rangle \\
    = \langle v_n^2 v_m^2 \rangle
      - \langle v_n^2 \rangle \langle v_m^2 \rangle .
  \end{multline}
  \par
}
}
{
{
  \tolerance=1000
  \emergencystretch=1em
  \begin{equation}
  \begin{aligned}
    \mathrm{SC}(n,m) &= 
      \langle\!\langle
        \cos(n\phi_1 + m\phi_2 - n\phi_3 - m\phi_4)
      \rangle\!\rangle \\
    &\quad - \langle\!\langle \cos[n(\phi_1 - \phi_2)] \rangle\!\rangle
      \langle\!\langle \cos[m(\phi_1 - \phi_2)] \rangle\!\rangle \\
    &= \langle v_n^2 v_m^2 \rangle
      - \langle v_n^2 \rangle \langle v_m^2 \rangle .
  \end{aligned}
  \end{equation}
  \par
}
}
This technique, originally introduced in Ref.~\cite{ALICE:2016kpq}, follows the same framework as described in Ref.~\cite{Bilandzic:2013kga}.
The SC observable isolates correlations between fluctuations of flow harmonics beyond trivial 
statistical fluctuations. Similarly, symmetric cumulants of six particles [SC($k,l,m$)] 
involve combinations of the second orders of three different $v_n$ coefficients~\cite{Mordasini:2019hut}.

Higher-order mixed harmonic cumulants (MHC) involving six or eight particles, 
also called $m$-observable cumulants~\cite{Moravcova:2020wnf} and first studied in Ref.~\cite{ALICE:2021adw}, 
provide further suppression of nonflow correlations and probe the nonlinear hydrodynamic response and 
mode-coupling effects between different flow harmonics with enhanced sensitivity~\cite{Moravcova:2020wnf, ALICE:2021adw}. 
For example, the six-particle MHC involving $v_2^4$ and $v_3^2$ is defined as: 
\begin{equation}\label{eq:eq12}
\text{MHC}(v_{2}^{4},v_{3}^{2})=\langle v_{2}^{4}v_{3}^{2}\rangle-4\langle v_{2}^{2}v_{3}^{2}\rangle\langle v_{2}^{2}\rangle-\langle v_{2}^{4}\rangle\langle v_{3}^{2}\rangle+4\langle v_{2}^{2}\rangle^{2}\langle v_{3}^{2}\rangle . 
\end{equation}
By construction, an $m$-particle cumulant removes contributions from correlations 
involving fewer than $m$ particles, thereby reducing contamination 
from nonflow sources such as jets or resonance decays even further~\cite{Moravcova:2020wnf}. Similar cumulants can be defined for 
combinations of different orders of $v_2$ and $v_3$, and also $v_2$ and $v_4$, following the prescription in 
Ref.~\cite{ALICE:2021adw}.

To enable more direct comparisons between different collision systems and between data and theoretical calculations, normalized cumulants are used. The normalized four- and six-particle symmetric cumulants (NSC) are defined as:
\begin{equation} 
\mathrm{NSC}(n, m) = \frac{\mathrm{SC}(n, m)}{\langle v_{n}^{2} \rangle \langle v_{m}^{2} \rangle} \label{eq:nsc2}
\end{equation}
\begin{equation} 
\mathrm{NSC}(k, l, m) = \frac{\mathrm{SC}(k, l, m)}{\langle v_{k}^{2} \rangle \langle v_{l}^{2} \rangle \langle v_{m}^{2} \rangle }. \label{eq:nsc3}
\end{equation}
These normalizations improve the comparison between different collision systems and 
with theoretical predictions. They do so by removing dependencies on the magnitudes 
of the flow coefficients, which can depend, for example, on the average \pt.

The normalized MHC (nMHC) scale the MHC by the average flow coefficients as:
\begin{equation}
\text{nMHC}(v_{m}^{k},v_{n}^{l}) = \frac{\text{MHC}(v_{m}^{k},v_{n}^{l})}{\langle v_{m}^{k}\rangle \langle v_{n}^{l}\rangle} \label{eq:eq16}
\end{equation}
\begin{equation}
\text{nMHC}(v_{m}^{k},v_{n}^{l},v_{p}^{q}) = \frac{\text{MHC}(v_{m}^{k},v_{n}^{l},v_{p}^{q})}{\langle v_{m}^{k}\rangle \langle v_{n}^{l}\rangle \langle v_{p}^{q}\rangle}. \label{eq:eq17}
\end{equation}
{\tolerance=1000
All cumulants are calculated using a $\abs{\Delta\eta} > 2$ selection in both the numerator and denominator of the normalized cumulants, further reducing residual nonflow effects. The two--particle correlation method is retained for $v_{n}\{2, \abs{\Delta\eta} > 2\}$ because it provides improved statistical precision while maintaining the same large pseudorapidity separation. In this regime, short-range correlations are strongly suppressed, and the resulting $v_{n}\{2, \abs{\Delta\eta} > 2\}$ values are consistent with those obtained from the two--particle cumulant technique within the quoted systematic uncertainties. These small differences do not affect the conclusions found from the model–-data comparisons.
\par}

The full mathematical details and further discussion of these methods can be found in Refs.~\cite{CMS:2019cyz, Bilandzic:2010jr, ALICE:2021adw}.

\section{Systematic uncertainties}

Systematic uncertainties are evaluated from four main sources for both \XeXe and \PbPb collisions. 
Only the maximum relative uncertainties across all observables are quoted below, with the largest values 
found for measurements involving $v_{4}$. In the following, percentage uncertainties are listed for 
\XeXe collisions first, followed by the value for \PbPb collisions. 
First, the track selection criteria are varied by comparing results 
using tighter (impact parameter significance $<$ 2 standard deviations, relative \pt uncertainty $<$ 5\%) 
and looser (5 standard deviations and 15\%) selections, resulting in maximum uncertainties of 
6 and 5\%. Second, uncertainties from the primary vertex 
position are estimated by comparing results with vertex positions along the beam axis in the 
ranges $\abs{z_{\mathrm{vtx}}} < 3$\unit{cm} and $3 < \abs{z_{\mathrm{vtx}}} < 15$\unit{cm}, yielding uncertainties up to 
6 and 5\% in the centrality range 0--10\%. Third, 
centrality calibration uncertainties, assessed by varying event selection criteria~\cite{CMS:2019cyz}, 
contribute up to 6 and 3\% in peripheral events. 
Finally, discrepancies between original track distributions generated by the 
\HYDJET model~\cite{Lokhtin:2005px} and those obtained after full detector simulation, 
reconstruction, and efficiency corrections produce systematic uncertainties up to 9 and 
5\%. The centrality dependence of these uncertainties 
varies with the analysis method: in two-particle correlations, the largest uncertainties appear in 
central collisions (9\% for $v_{4}\{2, \abs{\Delta\eta} > 2\}$), while in multiparticle cumulants, 
the maximum relative uncertainty is about 
6\% in mid-central to peripheral collisions. Overall, the various $v_{n}$ coefficients show 
consistent systematic trends, with partial cancellation in the cumulant results due to 
correlations between numerator and denominator terms defined in Eqs.~\eqref{eq:nsc2} to ~\eqref{eq:eq17}.

\section{Description of the models}

The \textsc{IP-Glasma} initial-condition (IC) model is an event generator incorporating event-by-event geometric and subnucleonic quantum fluctuations through dependencies on the impact parameter and the longitudinal momentum fraction $x$, where $x$ denotes the fraction of a nucleon's momentum carried by a quark or gluon in a high energy scattering process~\cite{Schenke:2012wb}. This model captures the fluctuating color charge configurations in ultrarelativistic nuclear collisions. The model has been shown to accurately reproduce experimental features of odd flow harmonics, particularly $v_3$, which are known to be dominated by initial-state fluctuations~\cite{Schenke:2012wb}.

The full \textsc{IP-Glasma}+\MUSIC+\textsc{UrQMD} framework~\cite{McDonald:2017ayb} builds upon the fluctuating initial states from \textsc{IP-Glasma-IC} by incorporating hydrodynamic evolution using \MUSIC~\cite{Gale:2013da}, a second-order relativistic viscous hydrodynamic program. The \MUSIC simulation employs a fixed specific shear viscosity to entropy density ratio $\eta/s = 0.12$ and a temperature-dependent bulk viscosity $\zeta/s(T)$ for the QGP phase until hadronization. Afterward, resonance decays and hadronic rescattering are simulated with \textsc{UrQMD} v3.4~\cite{Bass:1998ca, Schenke:2020mbo} until kinetic freeze-out.

For this analysis, four sets of Woods--Saxon deformation parameters for \Xe and one set for \Pb  are used. These deformation parameters originate from both theoretical calculations and empirical studies, and were provided by the authors of Ref.~\cite{Schenke:2012wb}. They are intended to span a representative range of plausible nuclear deformations for the \Xe nuclei and are summarized in Table~\ref{tab:WSparams}.

\begin{table}[h!t]
\centering
\topcaption{Woods--Saxon parameters used to model the nuclear density distributions for the $^{129}$\Xe nucleus in four different 
deformation configurations and for the $^{208}$\Pb nucleus. The parameters $R_0$, $a_0$, $\beta_2$, and $\beta_4$ 
are defined in Eq.~\eqref{eq:Rtheta}.}
\label{tab:WSparams}
\begin{tabular}{ccccc}
Nucleus & $R_{0}$ (fm) & $a_{0}$ (fm) & $\beta_{2}$ & $\beta_{4}$ \\
\hline
\Xe (a) & 5.601 & 0.492 & 0.207 & $-$0.003 \\
\Xe (b) & 5.420 & 0.570 & 0.000 & $-$0.003 \\
\Xe (c) & 5.420 & 0.570 & 0.162 & $-$0.003 \\
\Xe (d) & 5.420 & 0.570 & 0.207 & $-$0.003 \\
\Pb     & 6.647 & 0.537 & 0.006 & 0.000 \\
\end{tabular}
\end{table}

The \textsc{TRENTo-IC} model~\cite{Moreland:2014oya} is an initial-state model that does not assume any specific mechanism for pre-equilibrium dynamics, or thermalization. It deposits entropy in proportion to the generalized (typically geometric) mean of the nuclear overlap density between the colliding nuclei. 
This model simulates only the initial-state eccentricities, $\varepsilon_{n}$, and 
does not consider any final-state effects like hydrodynamic evolution or hadronic scattering. 
Eccentricities for \XeXe collisions are found using only two sets of deformation parameters, \Xe (a) for deformed and \Xe (b) for spherical configurations.

\section{Results}

\subsection{Sensitivity to nuclear deformation of the \Xe nucleus}

\begin{figure*}[ht!]
\includegraphics[width=1.0\textwidth]{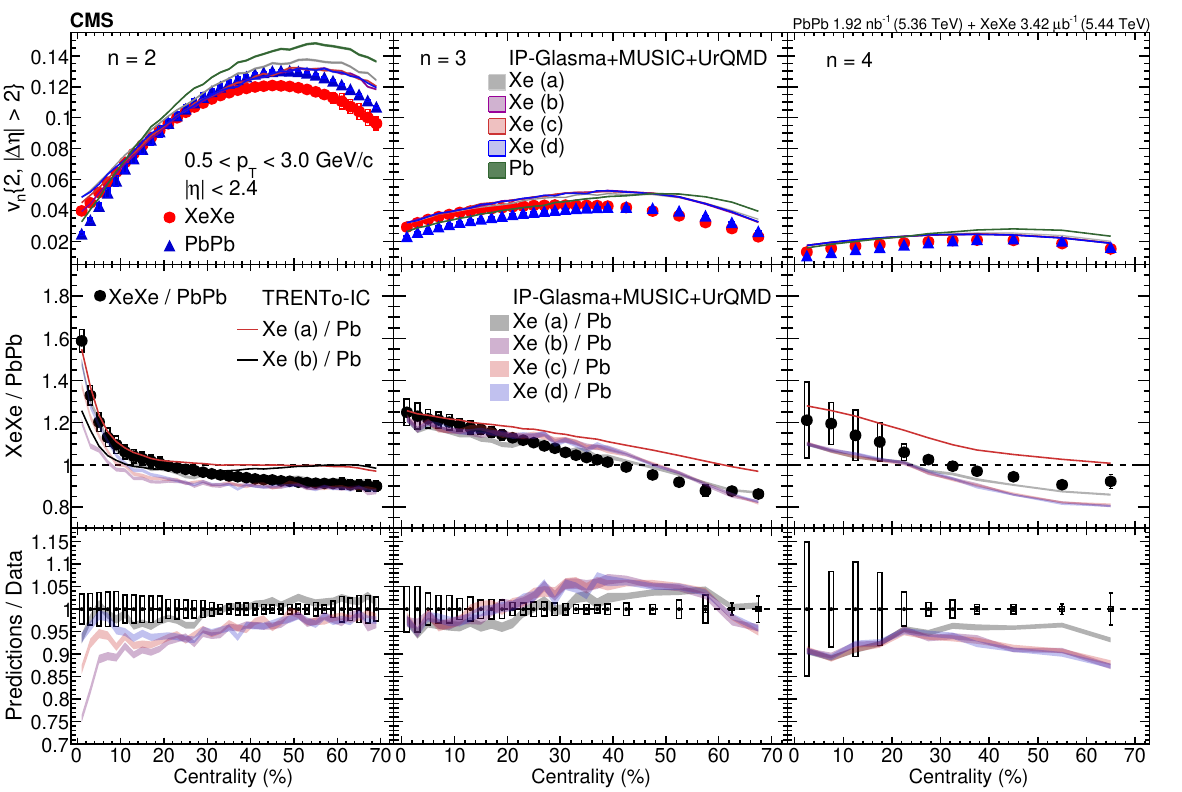}
\caption{The upper row shows $v_{n}\{2,\abs{\Delta\eta}>2\}$ for elliptic ($n = 2$, left), 
triangular ($n = 3$, middle), and quadrangular ($n = 4$, right) flow, as a function of centrality 
in \XeXe and \PbPb collisions, 
calculated using the 2-particle correlation method. The vertical bars (mostly smaller than the marker size) 
and the open boxes represent the statistical and systematic uncertainties, respectively. The shaded bands represent five 
different hydrodynamic predictions from 
the \textsc{IP-Glasma}+\MUSIC+\textsc{UrQMD} model using four \Xe parameter sets, and one \Pb parameter set; 
the band widths indicate the statistical uncertainties of the model calculations. Some model predictions in the upper row are 
partially obscured due to overlap with others. 
The middle row compares various theoretical predictions to the data for the \XeXe/\PbPb ratios. 
The red and black solid lines show predictions using ratios of the initial-state eccentricities 
$\varepsilon_{n}$ from the \textsc{TRENTo-IC} model 
with deformed (\Xe (a)) and spherical (\Xe (b)) parameters, respectively, 
both taken as ratios with respect to the \Pb  nucleus.  The lower row shows 
the ratios of the hydrodynamic predictions to data.}
\label{fig:vn2_plots}
\end{figure*}

Figure~\ref{fig:vn2_plots} presents the 2-particle correlation results for $v_{n}\{2,\abs{\Delta\eta}>2\}$ with $n=2,3,4$ as a function of centrality for \XeXe and \PbPb collisions. The upper, middle, and lower rows show the $v_{n}$ values, comparisons of model predictions to the data, and the ratios of the hydrodynamic model predictions to the data, respectively. 
All three flow harmonics exhibit larger values in \XeXe events than in \PbPb data for lower centrality percentages, which correspond to more central collisions, but smaller values in peripheral collisions. This enhancement of 
$v_2$ in central \XeXe collisions is likely driven by the quadrupole deformation of the \Xe nucleus, while the larger $v_3$ compared to \PbPb 
indicates stronger initial-state density fluctuations in \XeXe~\cite{Giacalone:2017dud}. 
In central collisions, enhanced values of $v_{4}$ in \XeXe relative to \PbPb collisions 
can be interpreted as arising from larger event-by-event fluctuations. In contrast, in mid-central and 
peripheral collisions, $v_{4}$ is dominated by nonlinear coupling 
to $v_{2}$~\cite{Giacalone:2017dud, Gardim:2011xv}. 
Since $v_{2}$ is smaller in \XeXe than in \PbPb in this centrality range, 
the resulting nonlinear contribution to $v_{4}$ is reduced in \XeXe, 
leading to \XeXe/\PbPb $v_{4}$ ratios below unity~\cite{Giacalone:2017dud}.

The middle row of Fig.~\ref{fig:vn2_plots} compares the \XeXe/\PbPb ratios to \textsc{IP-glasma}+\MUSIC+\textsc{UrQMD} and \textsc{TRENTo-IC} model predictions. 
The \textsc{IP-glasma}+\MUSIC+\textsc{UrQMD} hydrodynamic predictions are 
given for four \Xe nuclear shape parameter sets~\cite{Schenke:2012wb}, with statistical uncertainties shown as shaded bands. 
The \textsc{TRENTo-IC} model predicts the initial-state eccentricities $\varepsilon_n$ and their ratios.
The comparison to $v_n$ ratios assumes an approximately linear hydrodynamic response, 
$v_n \propto \varepsilon_n$, such that the flow ratios qualitatively reflect the corresponding eccentricity ratios. 
The predictions are displayed in Fig.~\ref{fig:vn2_plots} for $v_2$ using the \Xe (a) and (b) sets, 
but only with the \Xe (a) set for $v_3$ and $v_4$ due to their insensitivity to elliptical deformation. 
For a clearer comparison, the lower row 
in Fig.~\ref{fig:vn2_plots} displays the ratios of the hydrodynamic model predictions to data. 
Ratios are not shown for the \textsc{TRENTo-IC} model, as it provides initial-state eccentricities, 
while the ratios presented here involve final-state flow observables.

For the hydrodynamic model with parameter set (a), the largest value of the quadrupole deformation $\beta_2$ 
gives the best agreement for $v_2$ with a reduced $\chi^2/NDF \approx 0.99$, compared to values from 4.6 to 8.5 for the other 
sets. Set (d), having the same $\beta_2$ as set (a) but a larger value for the nuclear skin depth $a_0$, describes 
the data slightly better than sets (b) or (c) for the 0---10\% most central events ($\chi^2/NDF \approx 4.6$, as opposed to values 
of about 8.5 and 5.0 for sets (b) and (c), respectively). For $v_3$, set (a) also fits best ($\chi^2/NDF \approx 1.7$, 
compared to $\chi^2/NDF > 4$ for the others) for the 20--70\% centrality region, but is comparable or even slightly worse in 
other centrality regions. In the case of $v_{4}$, all 4 sets give comparable predictions for the 0--20\% centrality region, 
but set (a) performs best in the 20--70\% centrality region ($\chi^2/NDF \approx 6.8$ overall, compared to $\chi^2/NDF > 20$ for the others).

The \textsc{TRENTo-IC} model gives a poor description of $v_{2}$ for peripheral events (beyond about 35--40\% centrality) 
and fails to match $v_{3}$ and $v_{4}$ overall (except for extremely central collisions). These discrepancies are likely due 
to viscous damping effects, which are stronger in the smaller \XeXe system~\cite{Giacalone:2017dud}. However, 
for $v_{2}$ in the 0--20\% centrality region, which is more sensitive to deformation, the predictions using the 
deformed \Xe (a) parameters show better agreement with the data. In fact, this agreement is comparable to that 
for the hydrodynamic model using the same \Xe parameter set.

Differences between models and data may arise from multiple factors. Integrated flow harmonics can be over- or 
underestimated due to differences in the \pt ranges, affecting extraction of QGP transport coefficients 
[$\eta/s$, $\zeta/s(T)$]~\cite{Mantysaari:2024uwn}. The \textsc{IP-Glasma}+\MUSIC+\textsc{UrQMD} simulations 
use $\eta/s=0.12$, but theoretical uncertainties remain. The models also assume values of zero for both the octupole 
deformation ($\beta_3$) and triaxiality ($\gamma$) parameters, the latter describing deviations from axial symmetry. 
However, recent studies suggest that \Xe nuclei exhibit large deformation fluctuations due to a second-order phase transition 
along the \Xe isotope chain~\cite{Zhao:2024lpc}, which these models do not account for. Uncertainties in freeze-out criteria 
further impact predicted flow observables, contributing to discrepancies and uncertainties in extracting the 
\Xe deformation parameters. In addition, in peripheral collisions, contributions from non-thermalized 
or pp-like (core---corona) events may influence the measured correlations, effects which are not fully captured by the 
hydrodynamic models~\cite{Werner:2007zz}. While such contributions may partially cancel 
in the \XeXe/\PbPb ratios, some residual effects may remain due to differences in 
system size and geometry~\cite{Werner:2013tya}. 

\begin{figure*}[ht!]
\centering
\includegraphics[width=0.45\textwidth]{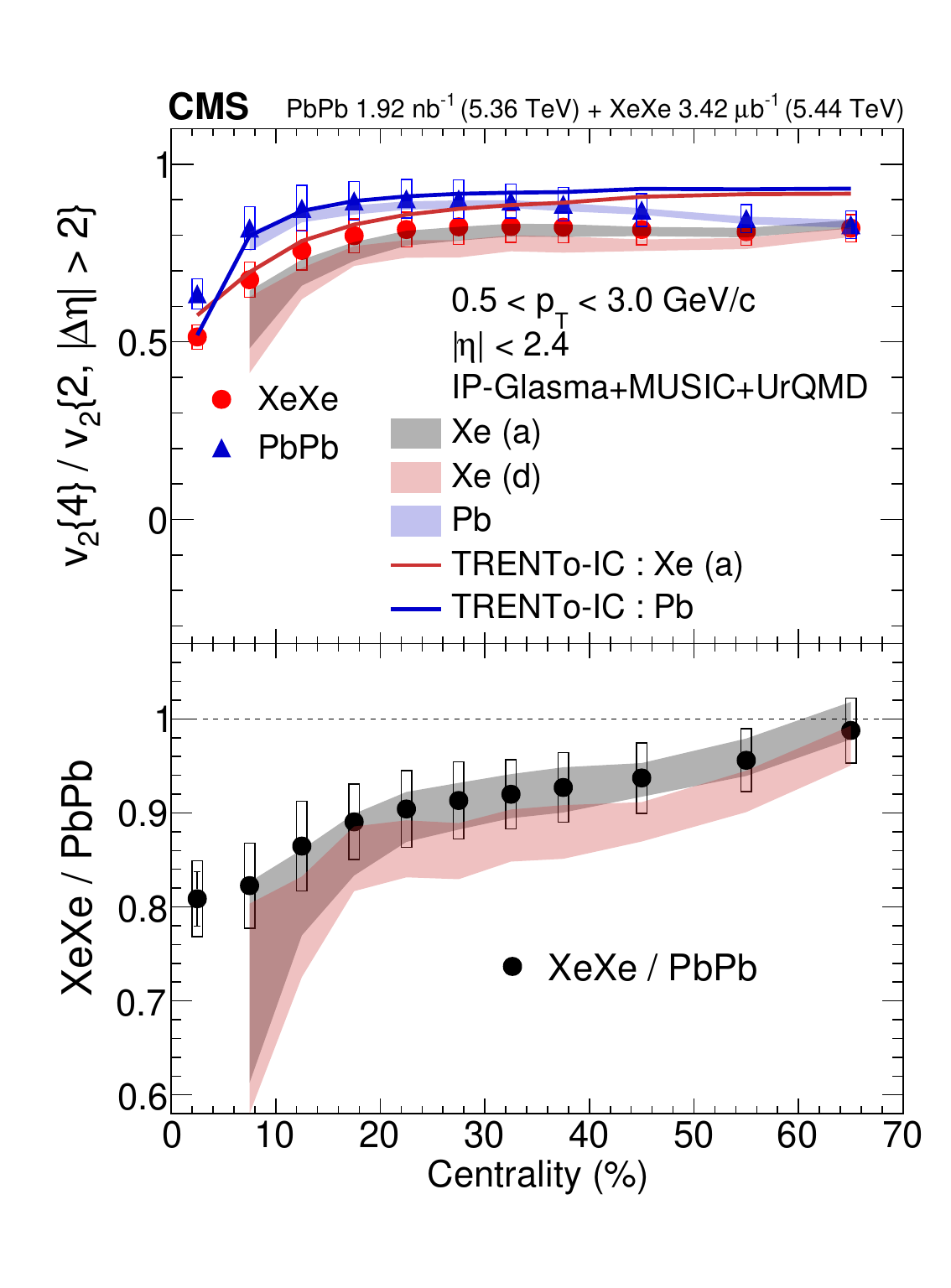}  
\includegraphics[width=0.45\textwidth]{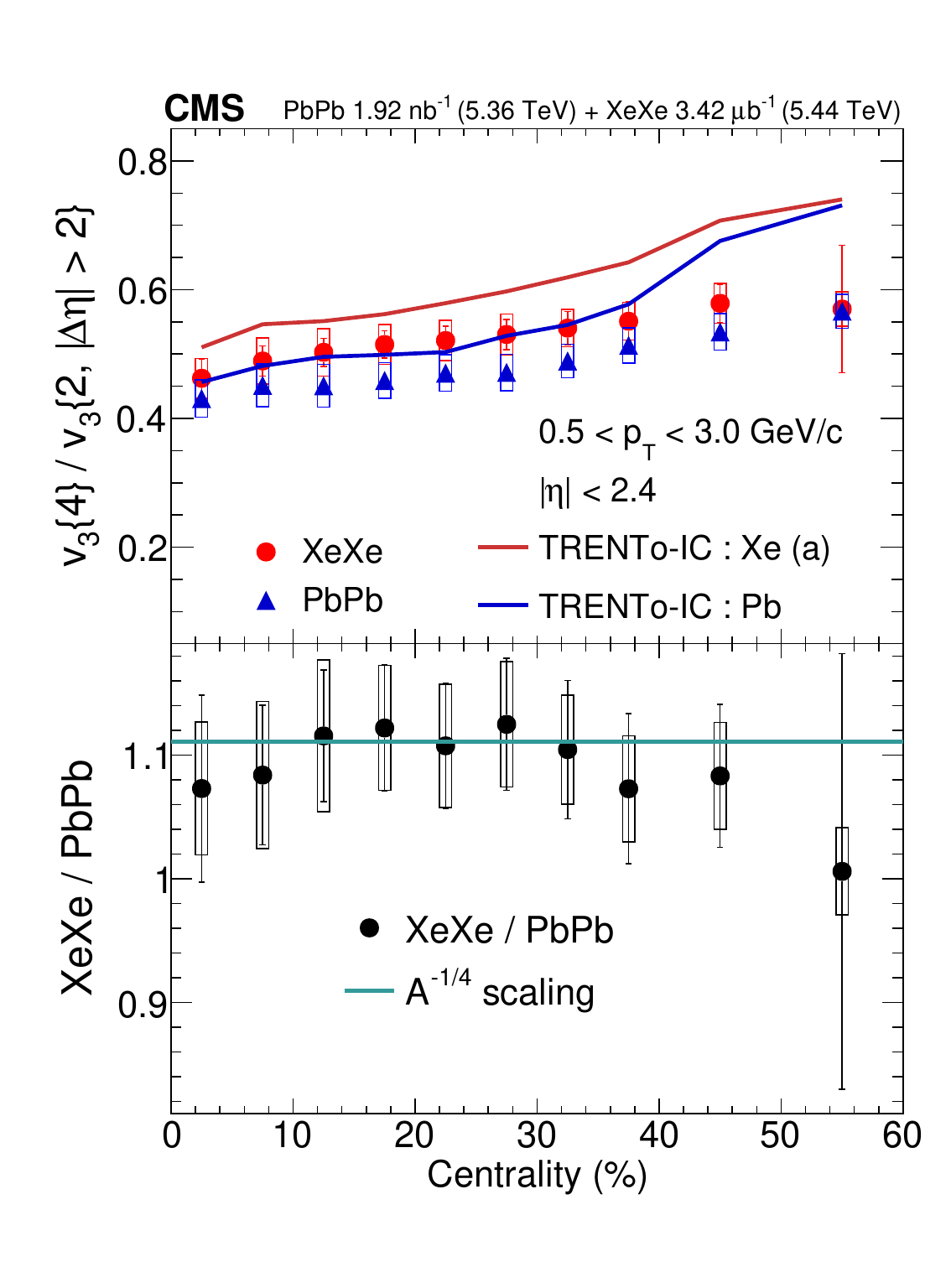}
\caption{
The upper panels show $v_{n}\{4\}/v_{n}\{2,|\Delta\eta|>2\}$ for elliptic ($n = 2$, left) 
and triangular ($n = 3$, right) flow, calculated from multiparticle cumulants, as a function of centrality in \XeXe and \PbPb collisions.
The vertical bars (mostly smaller than the marker size) on the dots and the open boxes represent the 
statistical and systematic uncertainties in the data, respectively.
The red and blue solid lines show the predictions from \textsc{TRENTo-IC} in terms of the initial-state eccentricities $\varepsilon_{n}$ 
for deformed \Xe (a) and \Pb nuclei.
The shaded bands show the hydrodynamic predictions from the \textsc{IP-Glasma}+\MUSIC+\textsc{UrQMD} model for \Xe (a) and \Xe (d); 
the band widths indicate the statistical uncertainties of the model calculations.
The lower panels display the ratio of the \XeXe to the \PbPb results.
The cyan line in the lower right panel shows the theoretical prediction for $v_{3}\{4\}/v_{3}\{2,\abs{\Delta\eta}>2\}$ based on $A^{-1/4}$ scaling.
}
\label{fig:vn4_plots}
\end{figure*}

The observable $v_{n}\{4\}/v_{n}\{2,\abs{\Delta\eta}>2\}$ provides direct information about 
event-by-event fluctuations of $v_{n}$~\cite{Giacalone:2017dud}. 
If $v_{n}$ fluctuates between events, this ratio is smaller than 1. 
The values of $v_{2}\{4\}/v_{2}\{2,\abs{\Delta\eta}>2\}$ exhibit a significant centrality dependence, 
as shown in the upper left panel of Fig.~\ref{fig:vn4_plots}, 
with the largest deviation from unity occurring in more central events. The systematically larger deviations from unity in \XeXe 
compared to \PbPb events indicate larger elliptic flow fluctuations in \XeXe collisions. 
These trends are also observed in both the initial-state \textsc{TRENTo-IC} and hydrodynamic \textsc{IP-Glasma}+\MUSIC+\textsc{UrQMD} predictions. 
In Ref.~\cite{Mantysaari:2024uwn}, this observable was shown to be sensitive to $a_{0}$, which characterizes the 
degree of nuclear diffuseness. A smaller $a_{0}$ corresponds to a sharper nuclear surface, which reduces 
event-by-event fluctuations of the initial elliptic geometry, particularly 
in semi-central collisions. As a result, the $v_{2}\{4\}/v_{2}\{2,\abs{\Delta\eta}>2\}$ ratio 
is less suppressed~\cite{Mantysaari:2024uwn}.

{\tolerance=1000
To further investigate this sensitivity, the double ratio of
$v_{2}\{4\}/v_{2}\{2,\abs{\Delta\eta}\!>\!2\}$ for \XeXe to \PbPb values is shown in the
lower left panel of Fig.~\ref{fig:vn4_plots}.
The theoretical predictions using parameter set \Xe (a) with
$a_{0} = 0.492$ more closely match the data than set (d) with $a_{0} = 0.570$,
reinforcing the conclusion based on $v_{2}\{2,\abs{\Delta\eta}\!>\!2\}$ alone.
\par}
In the following discussion, the \textsc{IP-Glasma}+\MUSIC+\textsc{UrQMD} predictions for \XeXe collisions will use only parameter set \Xe (a), 
which provides the closest match to data. 
For higher-order cumulants, all parameter sets remain compatible within the uncertainties.  

\subsection{Nonlinear hydrodynamic response through higher-order moments of flow harmonics}

The upper left panel in Fig.~\ref{fig:vn4_plots} also shows that the ratios $\varepsilon_{2}\{4\}/\varepsilon_{2}\{2\}$ from \textsc{TRENTo-IC} (solid lines) exceed the corresponding $v_{2}$ ratios in more peripheral collisions, indicating a nonlinear hydrodynamic response that increases towards peripheral collisions~\cite{Noronha-Hostler:2015dbi}. This effect is more pronounced in \XeXe collisions (starting from around 30\% centrality, as opposed to 50\% centrality in \PbPb collisions), suggesting stronger nonlinearity in the smaller \XeXe system. The same nonlinear response is also observed in the upper right panel of Fig.~\ref{fig:vn4_plots}, where the predictions from \textsc{TRENTo-IC} for $\varepsilon_{3}\{4\}/\varepsilon_{3}\{2\}$ (solid lines) exceed the corresponding $v_{3}$ ratios. In contrast to what was seen for $v_{2}$, the values of $v_{3}\{4\}/v_{3}\{2, \abs{\Delta\eta} > 2\}$ for both the predictions and the data are closer to unity for \XeXe than for \PbPb collisions. Finite-size corrections to the central limit theorem, which induce non-Gaussian fluctuations in the initial triangularity $\varepsilon_{3}$, were predicted to result in an $A^{-1/4}$ scaling of $v_{3}\{4\}/v_{3}\{2, \abs{\Delta\eta} > 2\}$ (where $A$ is the atomic mass number)~\cite{Giacalone:2017dud}. The non-Gaussian fluctuations scale with the number of participant nucleons, which roughly scales with $A$, leading to the observed scaling behavior. The lower right panel in Fig.~\ref{fig:vn4_plots} shows the \XeXe/\PbPb ratio, with the cyan line representing $(129/208)^{-1/4} \approx 1.13$, which agrees with the data within uncertainties.

The actual values of $v_{2}\{4\}$ and  $v_{3}\{4\}$, without normalization by $v_{n}\{2,\abs{\Delta\eta}>2\}$, are given in %
\ifthenelse{\boolean{cms@external}}
{~\ref{app_2}.}
{Appendix~\ref{app_2}.}

\subsubsection{Normalized symmetric cumulants involving two or three flow harmonics}

\begin{figure*}[ht!]
\centering
\includegraphics[width=0.45\textwidth]{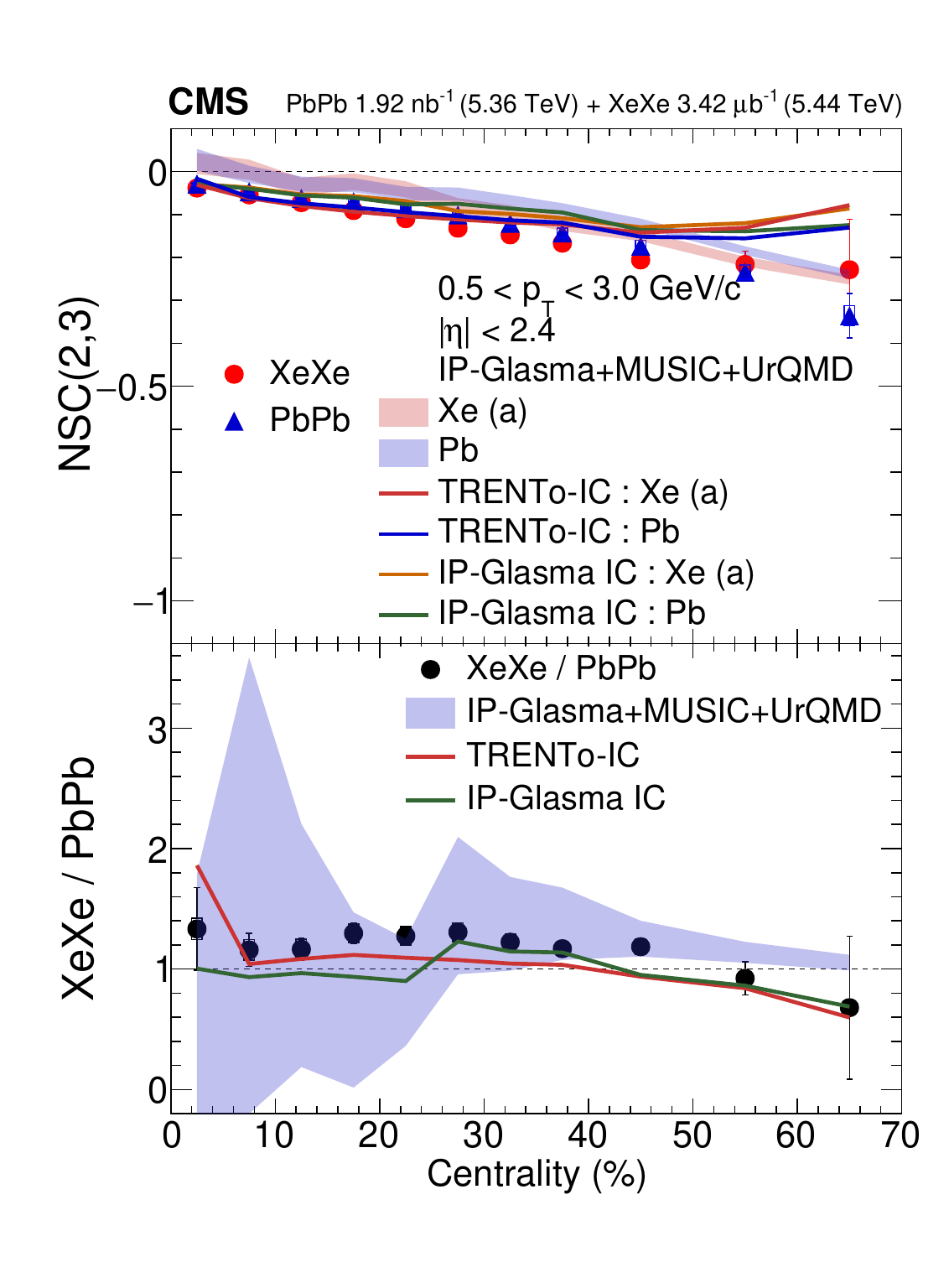} 
\includegraphics[width=0.45\textwidth]{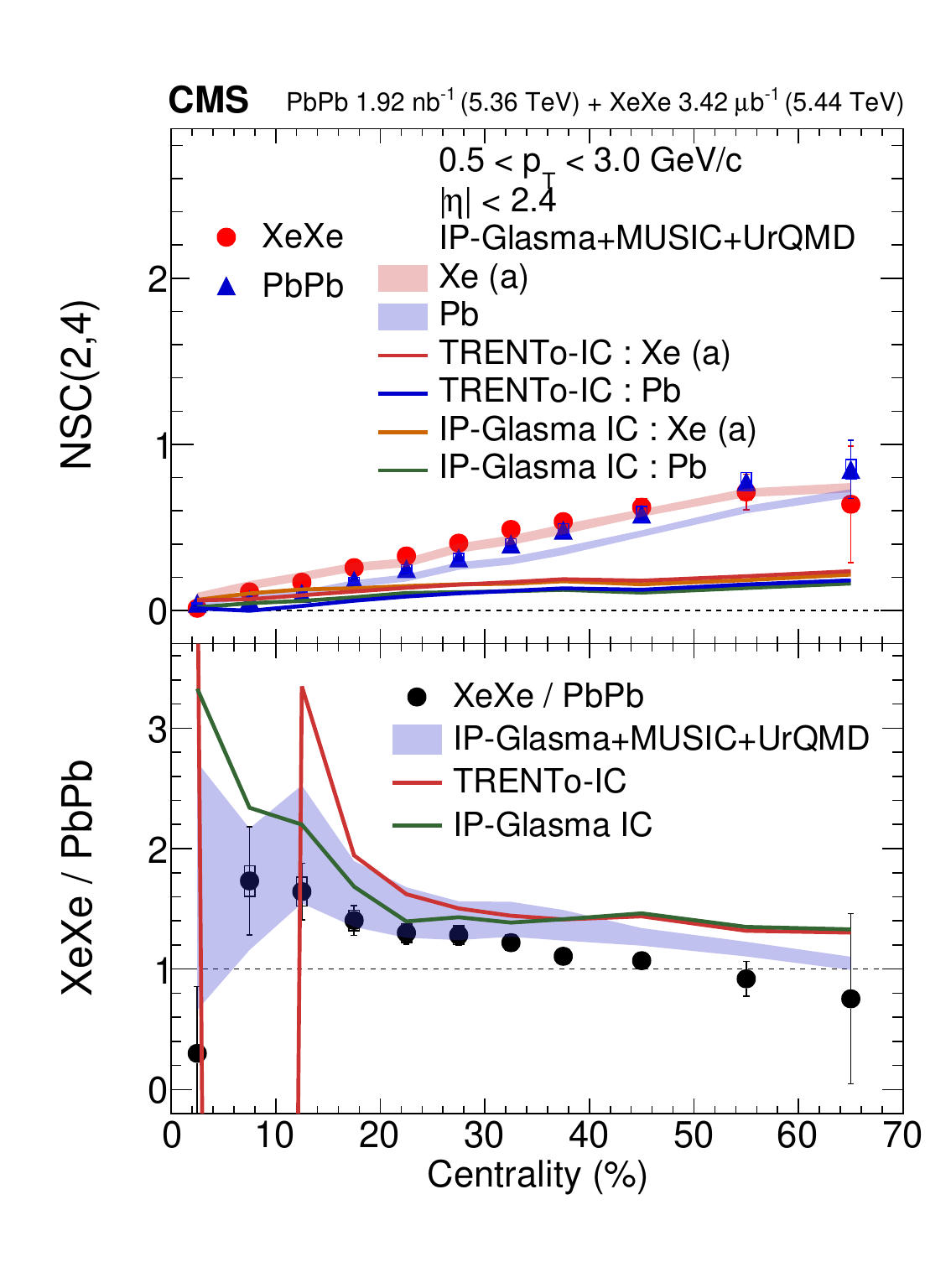}
\caption{The upper panels show the normalized symmetric cumulants NSC($2,3$) (left) and NSC($2,4$) (right) 
as functions of centrality in \XeXe and \PbPb collisions, calculated from multiparticle cumulants. 
The vertical bars and the open boxes represent the statistical and systematic uncertainties, respectively. 
The red and blue solid lines show the predictions from \textsc{TRENTo-IC} in terms of initial-state 
eccentricities $\varepsilon_{n}$ for deformed \Xe (\Xe (a)) and \Pb  nuclei. The orange and green solid lines show the predictions 
from the \textsc{IP-Glasma-IC} model with the same \Xe and \Pb  nucleus parameters. 
The shaded bands show the hydrodynamic predictions from the \textsc{IP-Glasma}+\MUSIC+\textsc{UrQMD} model; 
the band widths indicate the statistical uncertainties of the model calculations. 
The lower panels display the corresponding ratios of NSC($m,n$) for \XeXe/\PbPb. 
The red solid lines show the initial-state predictions from \textsc{TRENTo-IC}, while the green solid lines show those 
from \textsc{IP-Glasma-IC}. The blue shaded band is the hydrodynamic prediction from 
\textsc{IP-Glasma}+\MUSIC+\textsc{UrQMD}. In the 5--10\% centrality range, the \textsc{TRENTo-IC} NSC($2,4$) ratio 
is negative because the \PbPb prediction lies just below zero.}
\label{fig:nsc_plots}
\end{figure*}

Figure~\ref{fig:nsc_plots} presents the normalized four-particle symmetric cumulants involving two harmonics, NSC($2,3$) (left) and NSC($2,4$) (right), as functions of centrality in \XeXe and \PbPb collisions. In order to suppress nonflow effects, these are calculated using two particles from each of two separated $\eta$ regions: $-2.4 < \eta < -1$ and $1 < \eta < 2.4$. 
The NSC($2,3$) values probe the interplay between elliptic and triangular fluctuations influenced by eccentricities and deformation, while NSC($2,4$) is sensitive to nonlinear mode couplings and higher-multipole nuclear deformations (e.g., $\beta_4$), providing complementary constraints on the initial geometry. The results show that $v_{2}^{2}$ and $v_{3}^{2}$ are anticorrelated across all centralities, reflecting the anticorrelation between initial eccentricities $\varepsilon_{2}$ and $\varepsilon_{3}$, whereas $v_{2}^{2}$ and $v_{4}^{2}$ are positively correlated due to nonlinear mode coupling, where $v_{4}$ receives contributions proportional to $\varepsilon_{2}^{2}$~\cite{ALICE:2021adw}. The lower panels display the \XeXe/\PbPb ratios of NSC($2,3$) and NSC($2,4$), highlighting the differences between the initial states of the two systems. 
The \textsc{IP-Glasma}+\MUSIC+\textsc{UrQMD} hydrodynamic model captures the trends of both NSC($2,3$) and NSC($2,4$), 
albeit with a slight underestimation of the deviation from zero in some cases. 
Initial-state \textsc{TRENTo-IC} and \textsc{IP-Glasma}-IC models reproduce NSC($2,3$) except in peripheral collisions, 
where the linear relations $v_{2} \propto \varepsilon_{2}$ and $v_{3} \propto \varepsilon_{3}$ slightly 
break down~\cite{Noronha-Hostler:2015dbi}. For NSC(2,4), initial-state models increasingly underestimate the data from 
central to peripheral collisions, highlighting that these models lack the dynamical nonlinear mode-coupling generated 
during hydrodynamic expansion, through which $v_{4}$ receives a significant contribution proportional to $\varepsilon_{2}^{2}$.

For the \XeXe/\PbPb ratios shown in the lower row of Fig.~\ref{fig:nsc_plots}, 
\textsc{IP-Glasma}+\MUSIC+\textsc{UrQMD}, \textsc{IP-glasma-IC}, 
and \textsc{TRENTo-IC} all give roughly comparable agreement with the 
NSC($2,3$) data, with different models being closer to the data in different 
regions of centrality. Surprisingly, the ratios are well described even in 
the centrality regions where the two IC models fail to reproduce the 
individual \XeXe and \PbPb results. For the NSC($2,4$) ratios, 
the \textsc{IP-Glasma}+\MUSIC+\textsc{UrQMD} model gives a 
reasonable description of the data, while the two IC models 
(especially the \textsc{TRENTo-IC} model) do notably worse. 
As was the case for the NSC($2,3$) ratios, the NSC($2,4$) ratios from 
the IC models are much closer to the data than the individual \XeXe and \PbPb 
values. These comparisons highlight the role of NSC($2,3$) and NSC($2,4$) in 
constraining the initial-state geometry and its fluctuations, helping discriminate between 
different models of the initial entropy deposition and different sets of 
nuclear deformation parameters. This is because NSC($2,3$) is highly sensitive to correlations between the 
initial-state eccentricities $\varepsilon_2$ and $\varepsilon_3$, while NSC($2,4$), through its dependence on $v_4$, 
probes higher-order geometric structure and nonlinear mode-mixing effects. 
\begin{figure*}[ht!]
\centering
\includegraphics[width=0.45\textwidth]{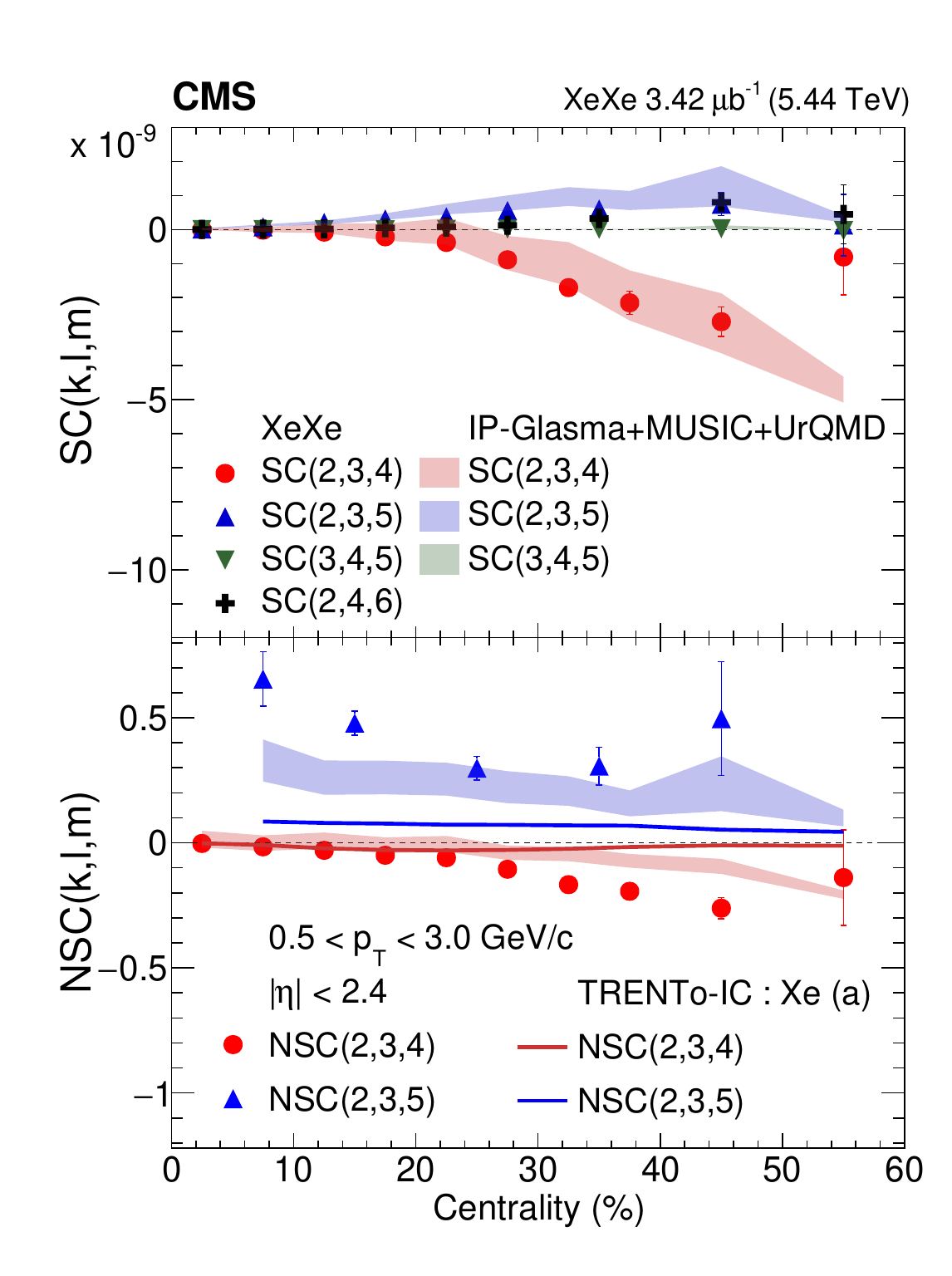} 
\includegraphics[width=0.45\textwidth]{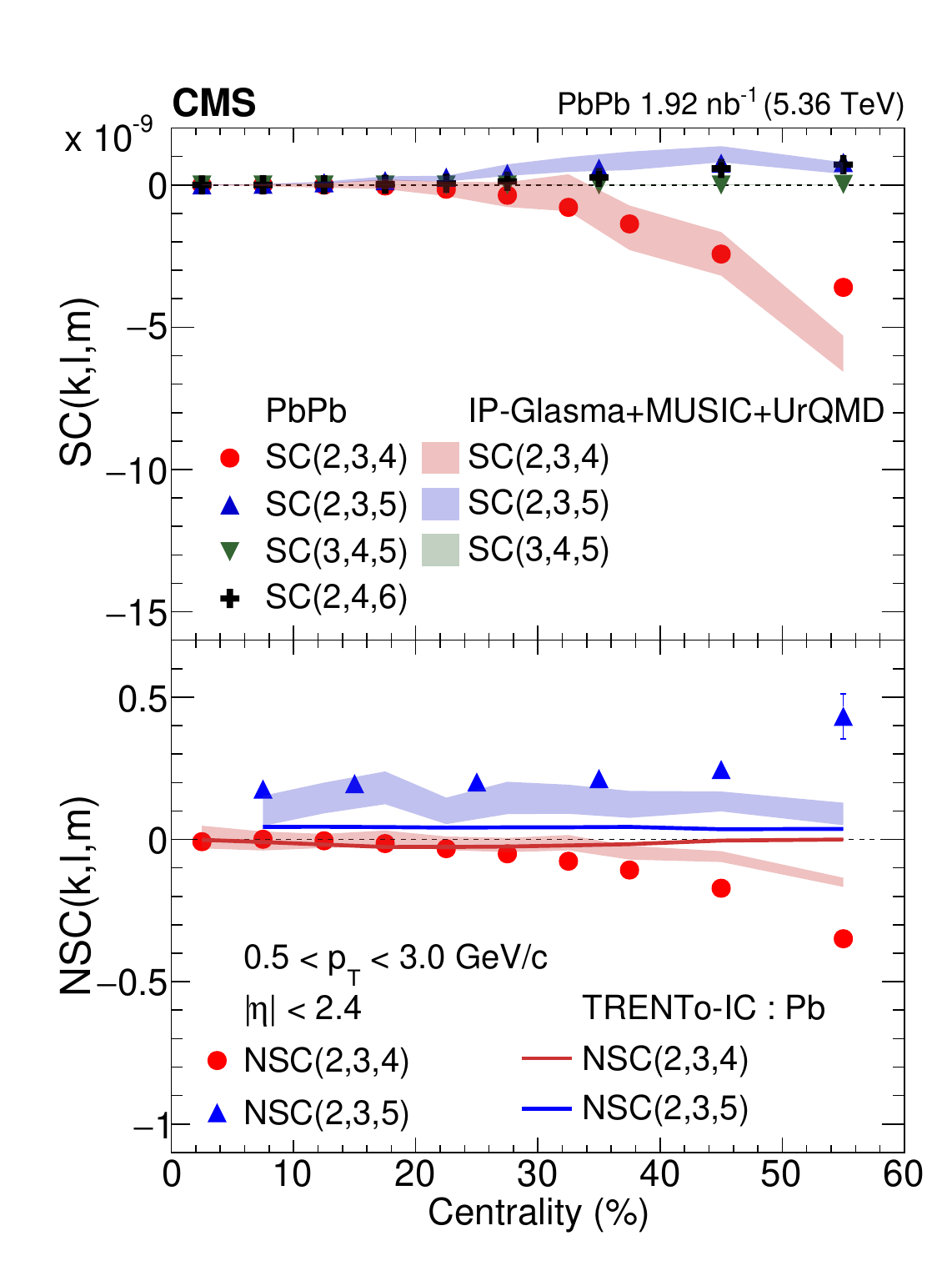}
\caption{A selection of six-particle three-harmonic SC and NSC cumulants as functions 
of centrality in \XeXe (left) and \PbPb (right) collisions, calculated from multiparticle cumulants. 
The vertical bars represent the statistical uncertainties, with the systematic uncertainties being 
negligible. The shaded bands show hydrodynamic predictions from the 
\textsc{IP-Glasma}+\MUSIC+\textsc{UrQMD} model for the \Xe (a) deformed nuclear shape 
parameters and for the spherical \Pb  nucleus; the band widths indicate the statistical uncertainties 
of the model calculations. The solid lines represent initial-state 
model predictions from \textsc{TRENTo-IC}.}
\label{fig:three_harmonic_cumulants}
\end{figure*}

Figure~\ref{fig:three_harmonic_cumulants} presents six-particle symmetric cumulants SC($k,l,m$) and their 
normalized counterparts NSC($k,l,m$) versus centrality for \XeXe collisions (left) measured 
for the first time, and \PbPb collisions (right). The upper rows display SC($2,3,4$), SC($2,3,5$), 
SC($3,4,5$), and SC($2,4,6$), while the lower rows show NSC($2,3,4$) and NSC($2,3,5$) in both systems. 
These measurements reveal genuine correlations among three different flow harmonics. In particular, 
the nonzero NSC($2,3,4$) values indicate fluctuations in the elliptical shape of the nuclear overlap 
region, beyond mere magnitude fluctuations of elliptic deformation~\cite{Mordasini:2019hut}. 
The NSC($2,3,4$) values in \XeXe and \PbPb are similar in the most central collisions, while in 
the mid-central range (15--45\%) the \XeXe values are systematically larger in magnitude, suggesting 
moderately stronger shape fluctuations in the initial geometry. The SC($3,4,5$) values are 
consistent with zero in both systems, as seen in Refs.~\cite{Mordasini:2019hut, ALICE:2021klf}, 
while the other SC($k,l,m$) cumulants have finite values. Notably, SC($2,3,5$) and NSC($2,3,5$) 
show different trends in \XeXe collisions, possibly 
indicating more nonlinear contribution for $v_5$ from $v_2$ and $v_3$ in the smaller system, 
with an even larger difference for the NSC data~\cite{ALICE:2021klf}. 

Hydrodynamic \textsc{IP-Glasma}+\MUSIC+\textsc{UrQMD} calculations provide an overall good 
description of the individual SC($k,l,m$) observables, 
but tend to underestimate the values of NSC($k,l,m$) in certain centrality intervals. 
In contrast, the initial-state \textsc{TRENTo-IC} predictions systematically underestimate 
NSC($k,l,m$), with the discrepancy becoming more pronounced toward peripheral collisions. 
As for the four-particle symmetric cumulants, this behaviour observed in \textsc{TRENTo-IC}
predictions is again attributed 
to the dynamical development of multiharmonic 
correlations during the hydrodynamic evolution of the medium, in particular due to 
nonlinear hydrodynamic contributions to $v_{4}$ and $v_{5}$ arising from 
$v_{2}$ and $v_{3}$~\cite{ALICE:2021klf, ALICE:2021adw}.

\subsubsection{Correlations between higher-order moments of two flow harmonics}

\begin{figure*}[ht!]
\centering
\includegraphics[width=0.45\textwidth]{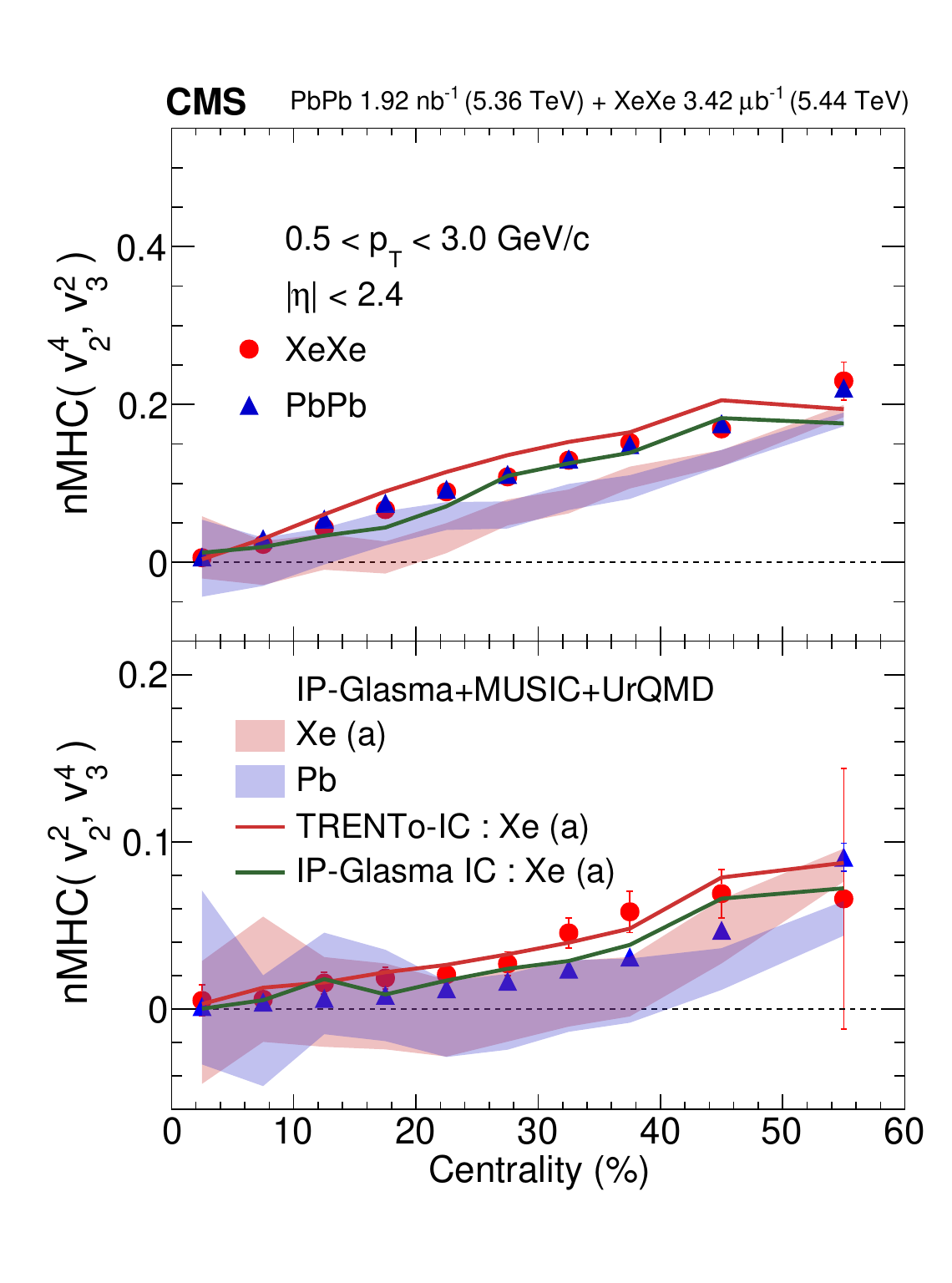} 
\includegraphics[width=0.45\textwidth]{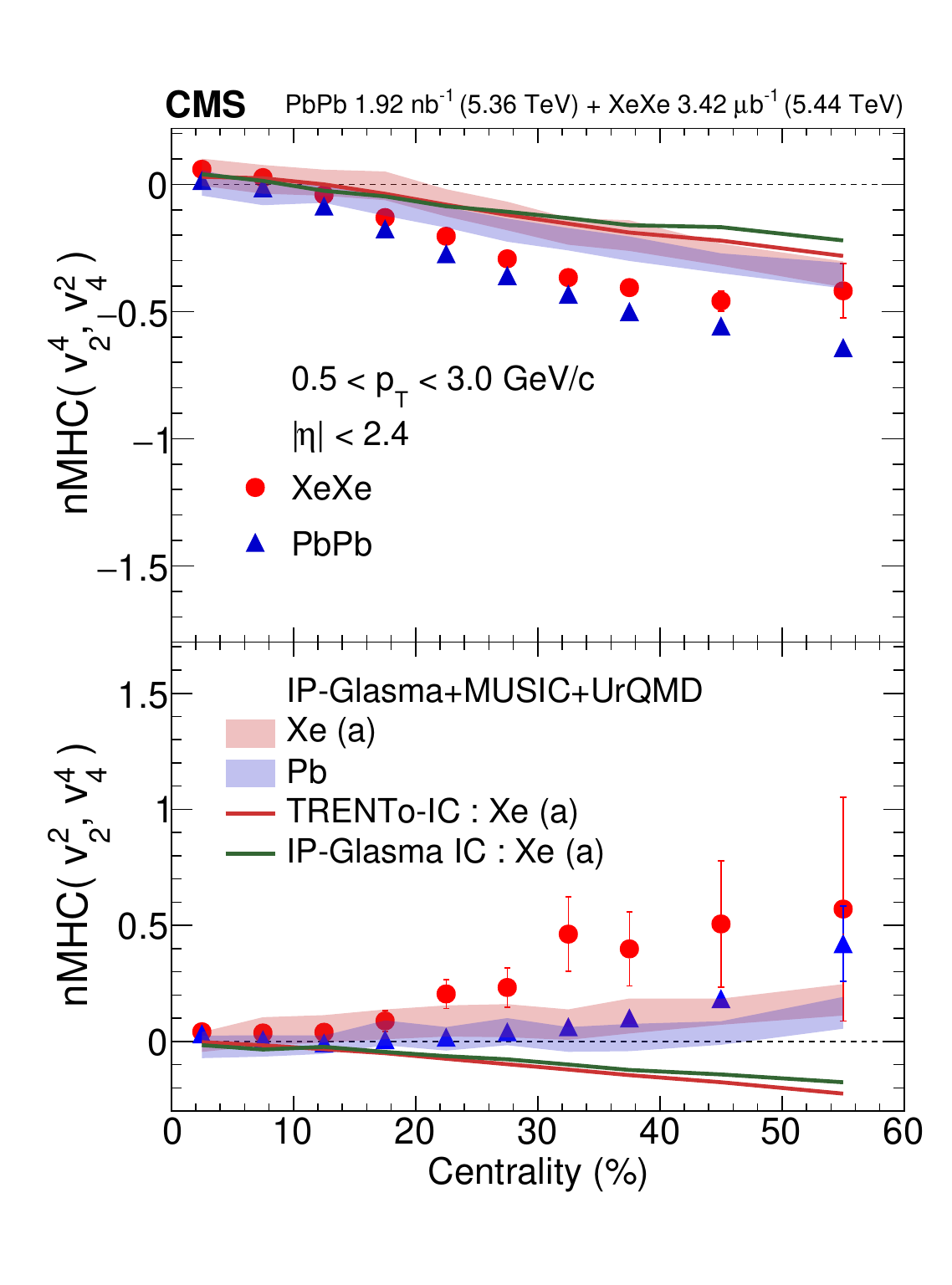}
\caption{Six-particle normalized mixed harmonic cumulants (nMHC) as functions of centrality in \XeXe and \PbPb collisions, calculated 
from multiparticle cumulants. Upper panels: nMHC($v_2^4, v_3^2$) (left) and nMHC($v_2^4, v_4^2$) (right). 
Lower panels: nMHC($v_2^2, v_3^4$) (left) and nMHC($v_2^2, v_4^4$) (right). The vertical bars represent the 
statistical uncertainties and the systematic uncertainties are negligible. The shaded bands show hydrodynamic predictions from the 
\textsc{IP-Glasma}+\MUSIC+\textsc{UrQMD} model for deformed \Xe (\Xe (a)) and \Pb  nuclei; the band widths 
indicate the statistical uncertainties of the model calculations. The solid lines represent initial-state model predictions.}
\label{fig:nmhc_centrality}
\end{figure*}

\begin{figure*}[ht!]
\centering
\includegraphics[width=0.45\textwidth]{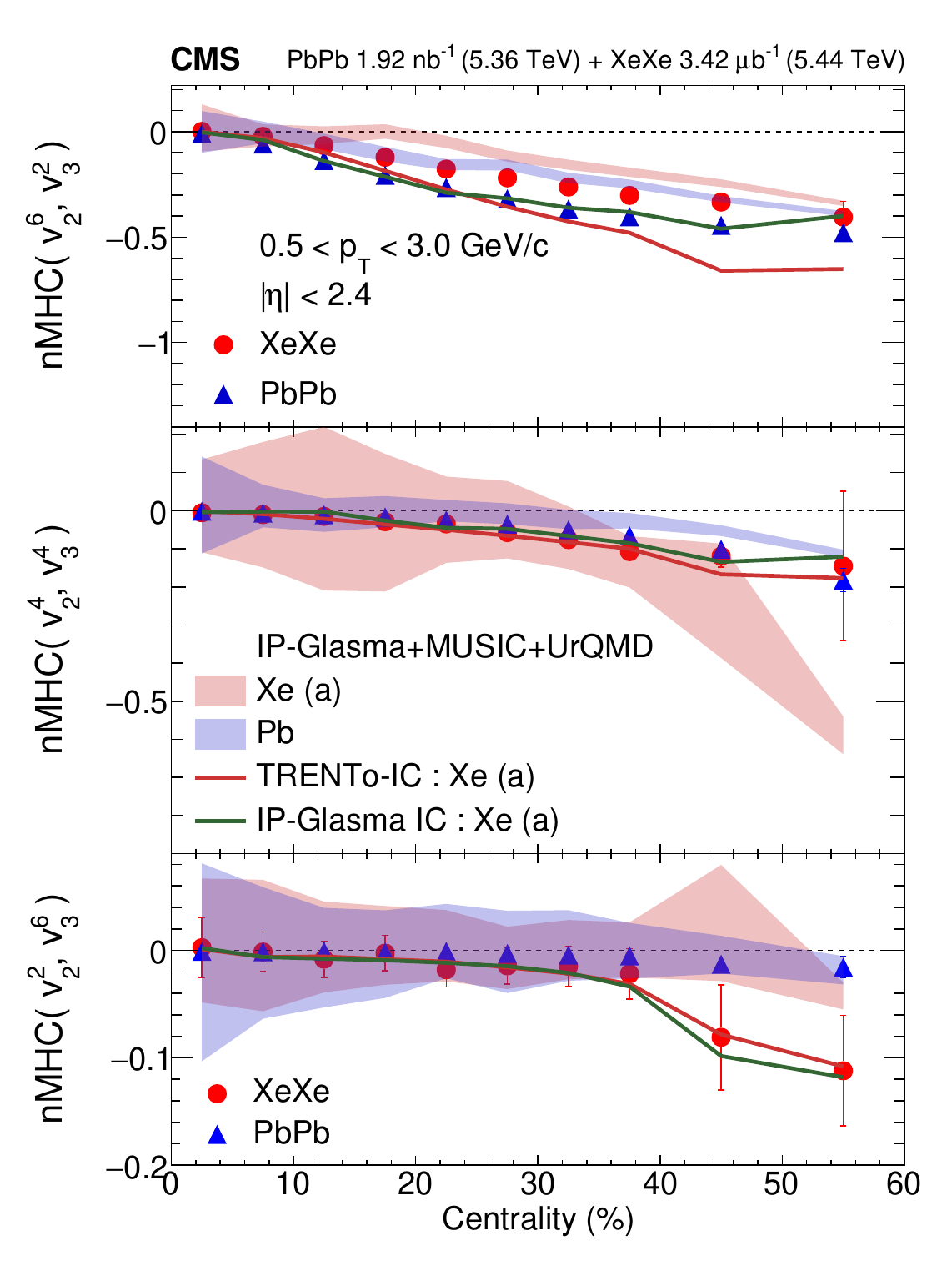} 
\includegraphics[width=0.45\textwidth]{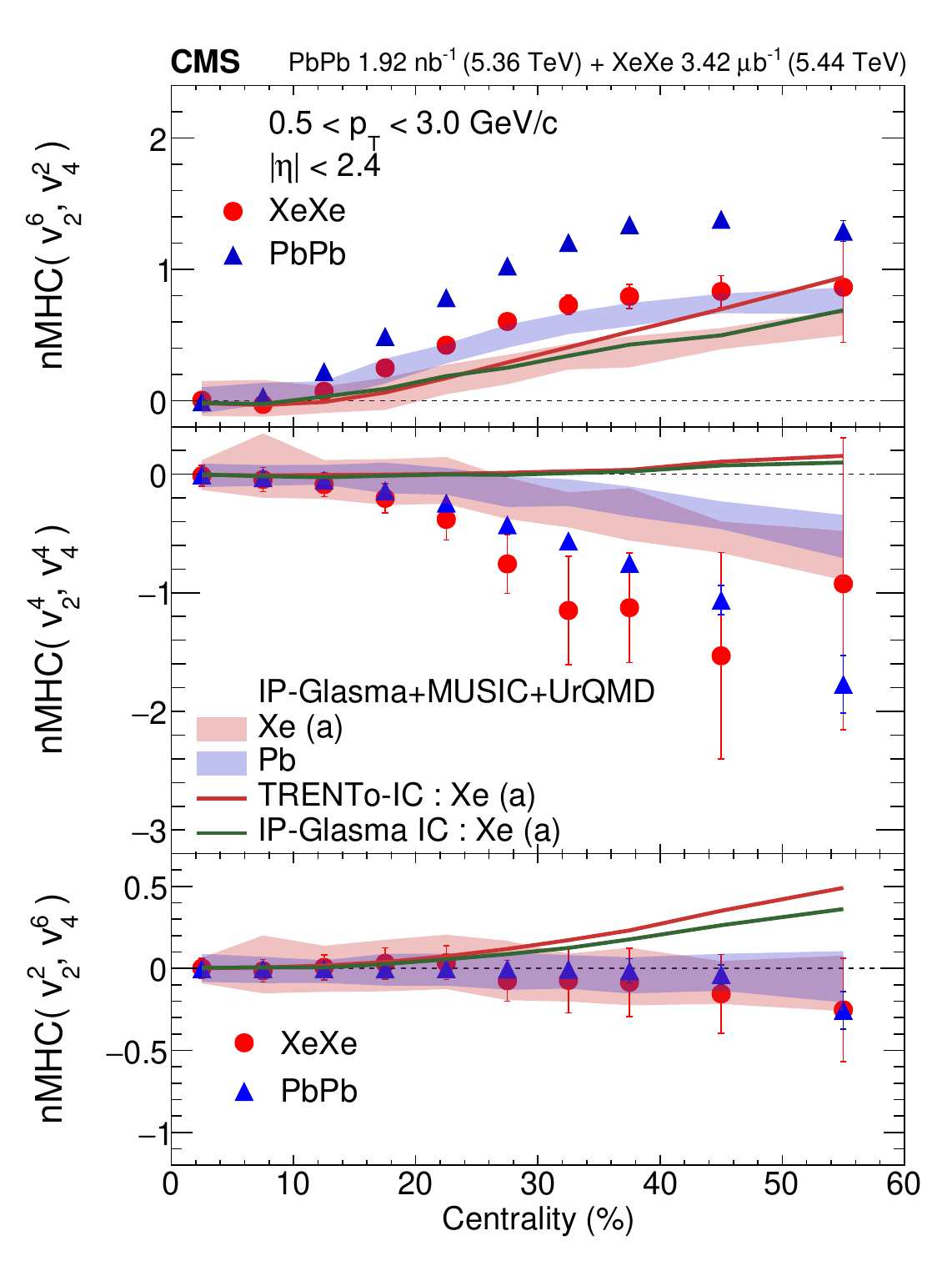}
\caption{Eight-particle normalized mixed harmonic cumulants (nMHC) as functions of centrality in \XeXe and \PbPb collisions, calculated from multiparticle cumulants. 
Left panels: nMHC($v_2^6, v_3^2$) (upper), nMHC($v_2^4, v_3^4$) (middle), and nMHC($v_2^2, v_3^6$) (lower). 
Right panels: nMHC($v_2^6, v_4^2$) (upper), nMHC($v_2^4, v_4^4$) (middle), and nMHC($v_2^2, v_4^6$) (lower). 
The vertical bars represent statistical uncertainties and the systematic uncertainties are negligible. 
The shaded bands show hydrodynamic predictions from the \textsc{IP-Glasma}+\MUSIC+\textsc{UrQMD} model for deformed \Xe (\Xe (a)) and  
\Pb nuclei; the band widths indicate the statistical uncertainties of the model calculations. The solid lines represent initial-state model predictions.}
\label{fig:8particle_nmhc}
\end{figure*}

Figures~\ref{fig:nmhc_centrality} and~\ref{fig:8particle_nmhc} present the first comparisons of $\text{nMHC}(v_{2}^{k}, v_{3}^{l})$ and $\text{nMHC}(v_{2}^{k}, v_{4}^{l})$ from \XeXe and \PbPb collisions for $k, l = 2, 4, 6$. The specific combinations are indicated in each panel. A key observation is that the magnitudes of $\text{nMHC}(v_{2}^{k}, v_{4}^{l})$ are larger than the corresponding values of $\text{nMHC}(v_{2}^{k}, v_{3}^{l})$ for the same $k$ and $l$ in both collision systems (note the differences in vertical scale between the left and right panels in both figures). This indicates a stronger correlation between higher order moments of $v_2$ and $v_4$, consistent with the trend observed for NSC($2,4$) versus NSC($2,3$). This can be attributed to the fact that both $v_2$ and $v_4$ are strongly influenced by the initial-state geometry.
A distinct pattern is observed in the magnitudes of correlations involving $v_3$:
\ifthenelse{\boolean{cms@external}}
{
\begin{multline}
\abs{\text{nMHC}(v_{2}^{6}, v_{3}^{2})} 
> \abs{\text{nMHC}(v_{2}^{4}, v_{3}^{2})} 
> \abs{\text{nMHC}(v_{2}^{4}, v_{3}^{4})} \\
> \abs{\text{nMHC}(v_{2}^{2}, v_{3}^{4})} 
> \abs{\text{nMHC}(v_{2}^{2}, v_{3}^{6})} ,
\label{eq:eq18}
\end{multline}
}
{
\begin{equation}
\abs{\text{nMHC}(v_{2}^{6}, v_{3}^{2})} 
> \abs{\text{nMHC}(v_{2}^{4}, v_{3}^{2})} 
> \abs{\text{nMHC}(v_{2}^{4}, v_{3}^{4})} 
> \abs{\text{nMHC}(v_{2}^{2}, v_{3}^{4})} 
> \abs{\text{nMHC}(v_{2}^{2}, v_{3}^{6})} ,
\label{eq:eq18}
\end{equation}
}
similar to the findings in Ref.~\cite{ALICE:2021adw}.

For $v_4$, the sequence differs slightly, with the second and third terms swapped:
\ifthenelse{\boolean{cms@external}}
{
\begin{multline}
\abs{\text{nMHC}(v_{2}^{6}, v_{4}^{2})} 
> \abs{\text{nMHC}(v_{2}^{4}, v_{4}^{4})} 
> \abs{\text{nMHC}(v_{2}^{4}, v_{4}^{2})} \\
> \abs{\text{nMHC}(v_{2}^{2}, v_{4}^{4})} 
> \abs{\text{nMHC}(v_{2}^{2}, v_{4}^{6})}.
\label{eq:eq19}
\end{multline}
}
{
\begin{equation}\label{eq:eq19}
\abs{\text{nMHC}(v_{2}^{6}, v_{4}^{2})} 
> \abs{\text{nMHC}(v_{2}^{4}, v_{4}^{4})} 
> \abs{\text{nMHC}(v_{2}^{4}, v_{4}^{2})} 
> \abs{\text{nMHC}(v_{2}^{2}, v_{4}^{4})} 
> \abs{\text{nMHC}(v_{2}^{2}, v_{4}^{6})}.
\end{equation}
}
Since $v_2$ and $v_4$ are strongly correlated, the inequality $\abs{\text{nMHC}(v_{2}^{4}, v_{4}^{4})} > \abs{\text{nMHC}(v_{2}^{4}, v_{4}^{2})}$ suggests that mode couplings involving equal orders of $v_2$ and $v_4$ exhibit stronger correlations.

System-size effects show a consistent trend across different $\text{nMHC}(v_2^k, v_3^l)$ and $\text{nMHC}(v_2^k, v_4^l)$ combinations. Given that initial-state fluctuations are expected to be stronger in the smaller \XeXe system, correlations involving higher-order moments of $v_3$ and $v_4$ ($v_3^4$, $v_3^6$, $v_4^4$, $v_4^6$) correlated with lower-order moments of $v_2$ ($v_2^2$) show enhanced values in \XeXe data, reflecting their greater sensitivity to fluctuations.

For correlations between $v_2$ and $v_3$, the four-, \mbox{six-,} and eight-particle cumulants are found to have negative, positive, and negative correlations, respectively, in both \XeXe and \PbPb collisions, in qualitative agreement with initial-state \textsc{TRENTo-IC} predictions~\cite{Zhao:2017yhj}. For example, comparing the upper left panels of Figs.~\ref{fig:nsc_plots},~\ref{fig:nmhc_centrality}, and~\ref{fig:8particle_nmhc}: NSC($2,3$) (4-particle), $\text{nMHC}(v_2^4, v_3^2)$ (6-particle), and $\text{nMHC}(v_2^6, v_3^2)$ (8-particle) follow this negative, positive, negative pattern.

However, this pattern does not hold for correlations between $v_2$ and $v_4$: correlations involving higher-order moments of $v_4$ ($l = 4, 6$) swap signs for the same order $n$-particle cumulant. For instance, $\text{nMHC}(v_2^4, v_4^2)$ and $\text{nMHC}(v_2^2, v_4^4)$ have negative and positive correlations, respectively, despite both being 6-particle cumulants.

{\tolerance=800
\emergencystretch=1em
The right hand panels in Figs.~\ref{fig:nmhc_centrality} and~\ref{fig:8particle_nmhc} compare the hydrodynamic and initial-state model predictions for correlations of various orders of $v_2$ and $v_4$, similar to previous studies using $v_2$ and $v_3$~\cite{ALICE:2021adw}. While the hydrodynamic models qualitatively reproduce all of the observed trends, initial-state models exhibit striking disagreements in the correlations of $\text{nMHC}(v_2^2, v_4^4)$, $\text{nMHC}(v_2^4, v_4^4)$, and $\text{nMHC}(v_2^2, v_4^6)$, predicting correlations opposite in sign to what is observed. This discrepancy likely arises because these observables depend on fourth- and sixth-order moments of $v_4$, which experience nonlinear hydrodynamic responses to $\varepsilon_2$ that grow towards peripheral collisions. This reinforces the conclusion that correlations between $v_2$ and $v_4$ predominantly develop dynamically during system evolution, 
rather than being fully determined by initial-state correlations. 
These correlations primarily arise from nonlinear hydrodynamic response and are 
influenced by transport properties such as shear and bulk viscosities~\cite{ALICE:2021adw}. 
\par}

Table~\ref{tab:correlationSigns} summarizes the signs of the correlations observed for different moments of $v_2$ and $v_3/v_4$ as functions of centrality.

\begin{table*}[ht!]
\centering
\topcaption{Signs of correlations between different moments of $v_2$ and $v_3/v_4$ as functions of centrality.}
\label{tab:correlationSigns}
\begin{tabular}{ccccc}
$n$-particle & Quantity & Correlation & Quantity & Correlation \\ 
cumulant & & sign & & sign \\ 
\hline
$4$ & NSC($2,3$) & $-$ & NSC($2,4$) & $+$ \\ 
$6$ & nMHC($v_2^4, v_3^2$) & $+$ & nMHC($v_2^4, v_4^2$) & $-$ \\ 
$6$ & nMHC($v_2^2, v_3^4$) & $+$ & nMHC($v_2^2, v_4^4$) & $+$ \\ 
$8$ & nMHC($v_2^6, v_3^2$) & $-$ & nMHC($v_2^6, v_4^2$) & $+$ \\
$8$ & nMHC($v_2^4, v_3^4$) & $-$ & nMHC($v_2^4, v_4^4$) & $-$ \\
$8$ & nMHC($v_2^2, v_3^6$) & $-$ & nMHC($v_2^2, v_4^6$) & $-$ (within uncertainties) \\ 
\end{tabular}
\end{table*}

\section{Summary}

For the first time, individual flow harmonics, two- and three-harmonic correlations, and 
higher-order mixed harmonic cumulants are measured and compared in 
detail for xenon-xenon (\XeXe) and lead-lead (\PbPb) collisions 
at $\sqrtsNN = 5.44$ and 5.36\TeV, respectively. Theoretical predictions of the 
\textsc{IP-Glasma}+\MUSIC+\textsc{UrQMD} hydrodynamic model, 
and the \textsc{TRENTo-IC} and \textsc{IP-Glasma-IC} initial-state models, 
calculated with various sets of deformation parameters for the $^{129}$\Xe nucleus, are compared 
to the data. The final-state model calculation from \textsc{IP-Glasma}+\MUSIC+\textsc{UrQMD} 
with nuclear deformation parameters $a_{0} = 0.492$ and $\beta_{2} = 0.207$ for 
$^{129}$\Xe provides the best agreement among tested parameter sets. Here, 
$a_{0}$ is the nuclear skin depth, and $\beta_{2}$ is the quadrupole deformation parameter. 
This comparison indicates that including a prolate deformation in the parameterization of the $^{129}$\Xe nuclear shape 
(in contrast to the doubly magic, nearly spherical $^{208}$\Pb nucleus) is necessary to reproduce 
the measured flow observables within the \textsc{IP-Glasma}+\MUSIC+\textsc{URQMD} framework.

On the other hand, the difference in initial-state predictions from the 
\textsc{TRENTo-IC} and \textsc{IP-Glasma-IC} models not only point to the sensitivity 
of these observables to the pre-equilibrium dynamics needed to model 
the experimental data, but also to the increasing nonlinearity of the higher-order 
moments of the flow harmonics in peripheral collisions. While differences between 
the two initial-state models are visible, a definitive conclusion about their relative performance cannot be drawn, 
as such a comparison without the inclusion of hydrodynamic evolution and freeze-out stages remains incomplete.

The comparison of higher-order moments in \XeXe and \PbPb 
collisions provides insight into the impact of system size and shape on 
these observables. Measurements involving higher-order moments of 
fluctuation-driven flow harmonics, such as $v_{3}$ and $v_{4}$, are, as expected, larger 
in the smaller \XeXe system.

Subtle differences exist between data and model calculations with the current sets of parameters. These 
differences highlight the importance of fine-tuning the model parameters, including not only the nuclear deformation 
parameters but also those related to hydrodynamic calculations, such as transport coefficients and freeze-out criteria. 
Such analyses can significantly improve the overall understanding of initial conditions in heavy ion collisions and 
transport properties of the quark-gluon plasma created at the LHC. Furthermore, this work 
provides a data-driven, multi-faceted probe of flow fluctuations, offering direct sensitivity 
to correlations arising from nonlinear hydrodynamic response.

\begin{acknowledgments}
We congratulate our colleagues in the CERN accelerator departments for the excellent performance of the LHC and thank the technical and administrative staffs at CERN and at other CMS institutes for their contributions to the success of the CMS effort. In addition, we gratefully acknowledge the computing centers and personnel of the Worldwide LHC Computing Grid and other centers for delivering so effectively the computing infrastructure essential to our analyses. Finally, we acknowledge the enduring support for the construction and operation of the LHC, the CMS detector, and the supporting computing infrastructure provided by the following funding agencies: SC (Armenia), BMBWF and FWF (Austria); FNRS and FWO (Belgium); CNPq, CAPES, FAPERJ, FAPERGS, and FAPESP (Brazil); MES and BNSF (Bulgaria); CERN; CAS, MoST, and NSFC (China); MINCIENCIAS (Colombia); MSES and CSF (Croatia); RIF (Cyprus); SENESCYT (Ecuador); ERC PRG, TARISTU24-TK10 and MoER TK202 (Estonia); Academy of Finland, MEC, and HIP (Finland); CEA and CNRS/IN2P3 (France); SRNSF (Georgia); BMFTR, DFG, and HGF (Germany); GSRI (Greece); NKFIH (Hungary); DAE and DST (India); IPM (Iran); SFI (Ireland); INFN (Italy); MSIT and NRF (Republic of Korea); MES (Latvia); LMTLT (Lithuania); MOE and UM (Malaysia); BUAP, CINVESTAV, CONACYT, LNS, SEP, and UASLP-FAI (Mexico); MOS (Montenegro); MBIE (New Zealand); PAEC (Pakistan); MES, NSC, and NAWA (Poland); FCT (Portugal); MESTD (Serbia); MICIU/AEI and PCTI (Spain); MOSTR (Sri Lanka); Swiss Funding Agencies (Switzerland); MST (Taipei); MHESI (Thailand); TUBITAK and TENMAK (T\"{u}rkiye); NASU (Ukraine); STFC (United Kingdom); DOE and NSF (USA).

\hyphenation{Rachada-pisek} Individuals have received support from the Marie-Curie program and the European Research Council and Horizon 2020 Grant, contract Nos.\ 675440, 724704, 752730, 758316, 765710, 824093, 101115353, 101002207, 101001205, and COST Action CA16108 (European Union); the Leventis Foundation; the Alfred P.\ Sloan Foundation; the Alexander von Humboldt Foundation; the Science Committee, project no. 22rl-037 (Armenia); the Fonds pour la Formation \`a la Recherche dans l'Industrie et dans l'Agriculture (FRIA) and Fonds voor Wetenschappelijk Onderzoek contract No. 1228724N (Belgium); the Beijing Municipal Science \& Technology Commission, No. Z191100007219010, the Fundamental Research Funds for the Central Universities, the Ministry of Science and Technology of China under Grant No. 2023YFA1605804, the Natural Science Foundation of China under Grant No. 12061141002, 12535004, and USTC Research Funds of the Double First-Class Initiative No.\ YD2030002017 (China); the Ministry of Education, Youth and Sports (MEYS) of the Czech Republic; the Shota Rustaveli National Science Foundation, grant FR-22-985 (Georgia); the Deutsche Forschungsgemeinschaft (DFG), among others, under Germany's Excellence Strategy -- EXC 2121 ``Quantum Universe" -- 390833306, and under project number 400140256 - GRK2497; the Hellenic Foundation for Research and Innovation (HFRI), Project Number 2288 (Greece); the Hungarian Academy of Sciences, the New National Excellence Program - \'UNKP, the NKFIH research grants K 131991, K 133046, K 138136, K 143460, K 143477, K 146913, K 146914, K 147048, 2020-2.2.1-ED-2021-00181, TKP2021-NKTA-64, and 2021-4.1.2-NEMZ\_KI-2024-00036 (Hungary); the Council of Science and Industrial Research, India; ICSC -- National Research Center for High Performance Computing, Big Data and Quantum Computing, FAIR -- Future Artificial Intelligence Research, and CUP I53D23001070006 (Mission 4 Component 1), funded by the NextGenerationEU program (Italy); the Latvian Council of Science; the Ministry of Education and Science, project no. 2022/WK/14, and the National Science Center, contracts Opus 2021/41/B/ST2/01369, 2021/43/B/ST2/01552, 2023/49/B/ST2/03273, and the NAWA contract BPN/PPO/2021/1/00011 (Poland); the Funda\c{c}\~ao para a Ci\^encia e a Tecnologia, grant CEECIND/01334/2018 (Portugal); the National Priorities Research Program by Qatar National Research Fund; MICIU/AEI/10.13039/501100011033, ERDF/EU, "European Union NextGenerationEU/PRTR", and Programa Severo Ochoa del Principado de Asturias (Spain); the Chulalongkorn Academic into Its 2nd Century Project Advancement Project, the National Science, Research and Innovation Fund program IND\_FF\_68\_369\_2300\_097, and the Program Management Unit for Human Resources \& Institutional Development, Research and Innovation, grant B39G680009 (Thailand); the Kavli Foundation; the Nvidia Corporation; the SuperMicro Corporation; the Welch Foundation, contract C-1845; and the Weston Havens Foundation (USA).
\end{acknowledgments}

\bibliography{auto_generated} 

\numberwithin{figure}{section}
\newpage
\appendix
\section{\texorpdfstring{$v_{2}\{4\}$ and $v_{3}\{4\}$}{v2\{4\} and v3\{4\}} \vs centrality}
\label{app_2}

\begin{figure}[ht!b] 
\centering 
\includegraphics[width=0.45\textwidth]{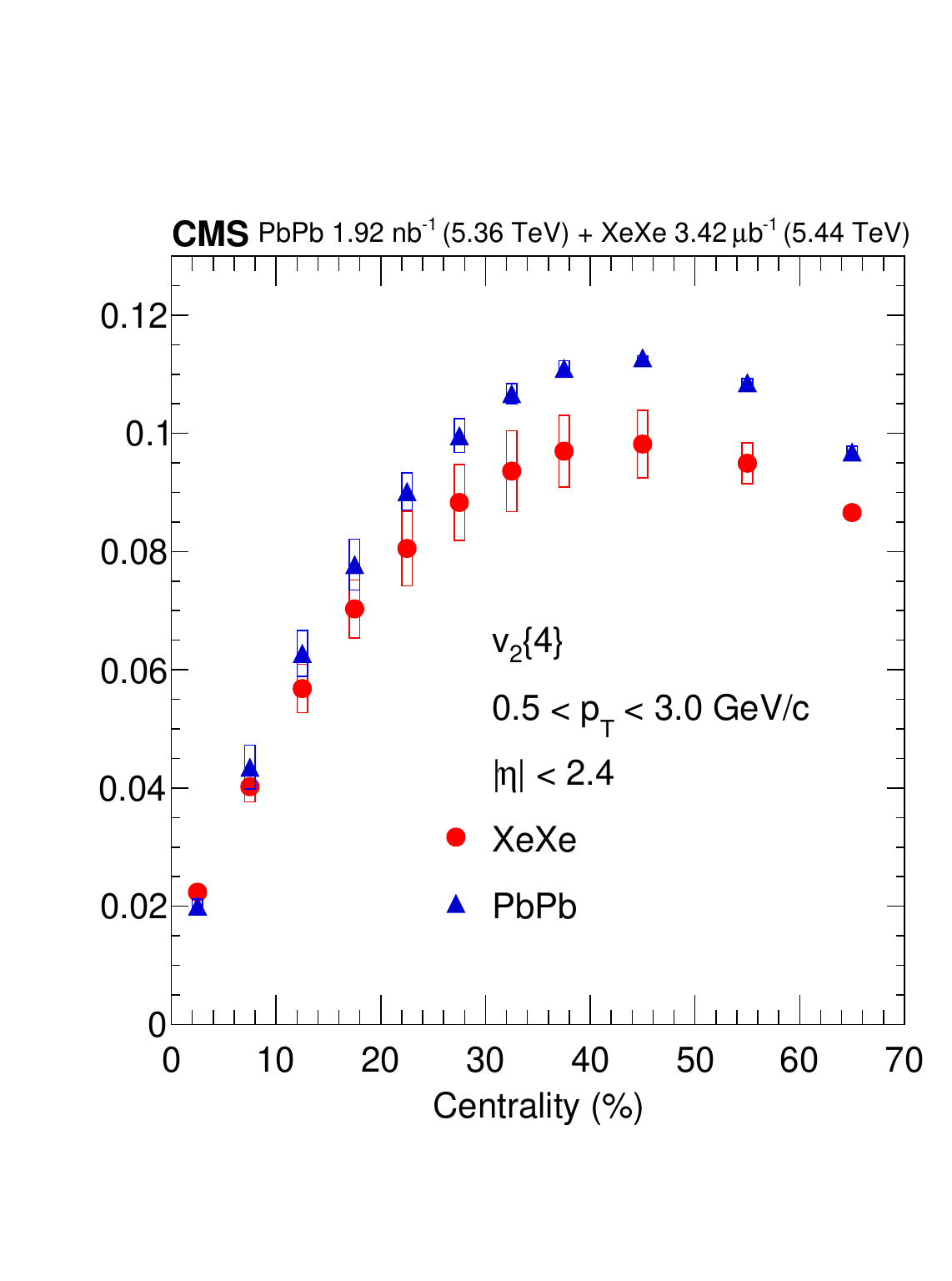} 
\includegraphics[width=0.45\textwidth]{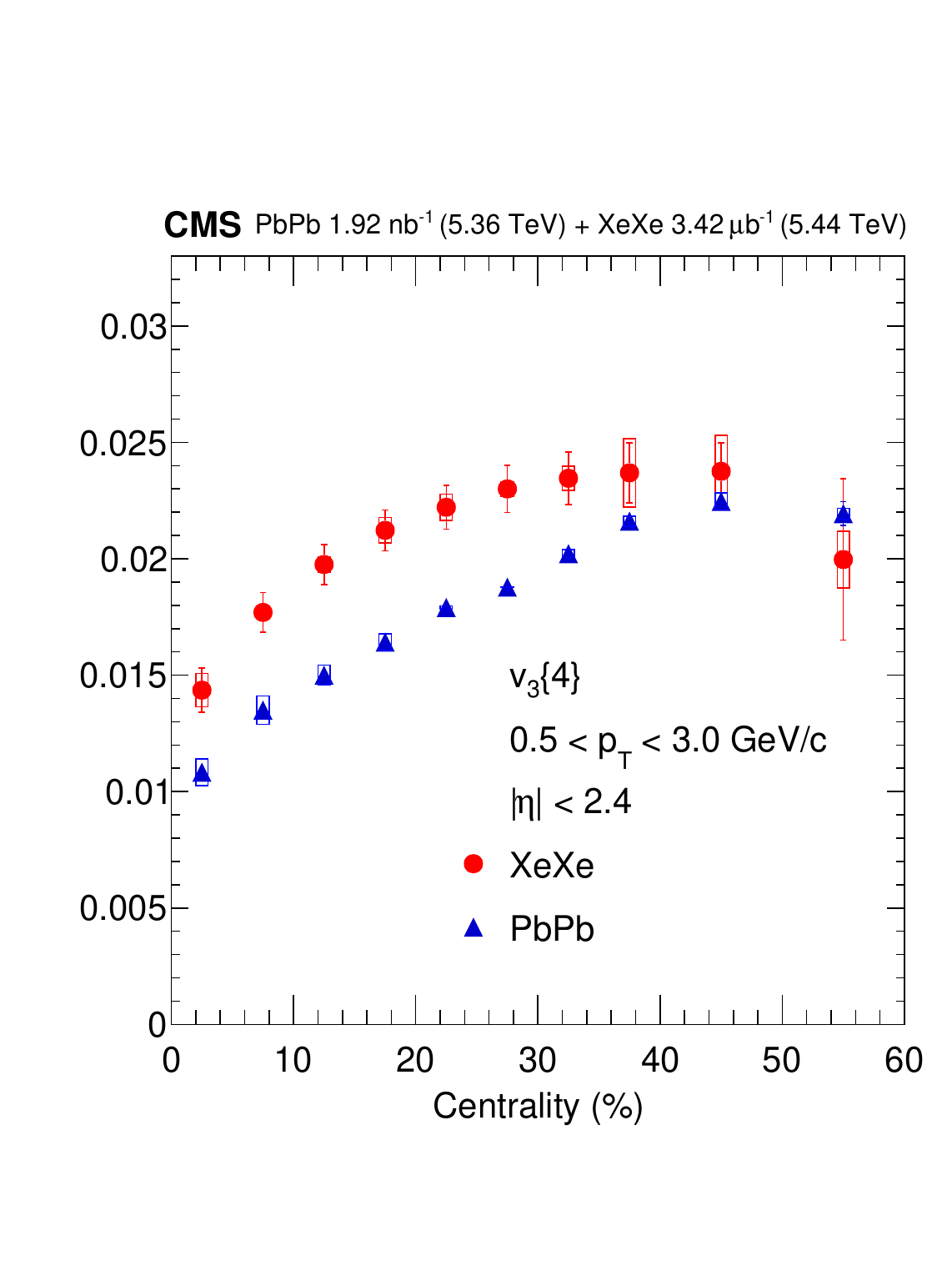} 
\caption{Measured $v_2\{4\}$ (\cmsLeft) and $v_3\{4\}$ (\cmsRight) cumulant values as a function 
of centrality for \XeXe and \PbPb collisions, calculated using 4-particle cumulants. The 
vertical bars and open boxes represent the statistical and systematic uncertainties, respectively.} 
\label{fig:v24_v34} 
\end{figure}

Figure~\ref{fig:v24_v34} shows similar measurements to those given in the upper panels of Fig.~\ref{fig:vn2_plots}, 
but now for $v_{2}\{4\}$ and $v_{3}\{4\}$. 

\cleardoublepage \section{The CMS Collaboration \label{app:collab}}\begin{sloppypar}\hyphenpenalty=5000\widowpenalty=500\clubpenalty=5000
\cmsinstitute{Yerevan Physics Institute, Yerevan, Armenia}
{\tolerance=6000
A.~Hayrapetyan, V.~Makarenko\cmsorcid{0000-0002-8406-8605}, A.~Tumasyan\cmsAuthorMark{1}\cmsorcid{0009-0000-0684-6742}
\par}
\cmsinstitute{Institut f\"{u}r Hochenergiephysik, Vienna, Austria}
{\tolerance=6000
W.~Adam\cmsorcid{0000-0001-9099-4341}, J.W.~Andrejkovic, L.~Benato\cmsorcid{0000-0001-5135-7489}, T.~Bergauer\cmsorcid{0000-0002-5786-0293}, M.~Dragicevic\cmsorcid{0000-0003-1967-6783}, C.~Giordano, P.S.~Hussain\cmsorcid{0000-0002-4825-5278}, M.~Jeitler\cmsAuthorMark{2}\cmsorcid{0000-0002-5141-9560}, N.~Krammer\cmsorcid{0000-0002-0548-0985}, A.~Li\cmsorcid{0000-0002-4547-116X}, D.~Liko\cmsorcid{0000-0002-3380-473X}, I.~Mikulec\cmsorcid{0000-0003-0385-2746}, J.~Schieck\cmsAuthorMark{2}\cmsorcid{0000-0002-1058-8093}, R.~Sch\"{o}fbeck\cmsAuthorMark{2}\cmsorcid{0000-0002-2332-8784}, D.~Schwarz\cmsorcid{0000-0002-3821-7331}, M.~Shooshtari\cmsorcid{0009-0004-8882-4887}, M.~Sonawane\cmsorcid{0000-0003-0510-7010}, W.~Waltenberger\cmsorcid{0000-0002-6215-7228}, C.-E.~Wulz\cmsAuthorMark{2}\cmsorcid{0000-0001-9226-5812}
% ...
\par}
\vskip\cmsinstskip
\dag:~Deceased\\
$^{1}$Also at Yerevan State University, Yerevan, Armenia\\
$^{2}$Also at TU Wien, Vienna, Austria\\
% ...
\end{sloppypar}
\end{document}